\documentclass[10pt]{article}
\usepackage{cite}
\usepackage{graphicx}
\usepackage{fullpage}

\newcommand{\PreserveBackslash}[1]{\let\temp=\\#1\let\\=\temp}
\newcommand{\Het}{$^3$He}
\newcommand{\Hef}{$^4$He}
\newcommand{\ncms}{$\rm\,cm^{-2}\,s^{-1}$}

\newcommand{\EMIT}{\textsc{emiT}}
\newcommand{\taun}{$\tau_{n}$}
\newcommand{\gv}{$g_V$}
\newcommand{\ga}{$g_A$}
\newcommand{\NPDG}{p(${\vec {\text n}}$,$\gamma$)d}
\newcommand{\NDTG}{d(${\vec {\text n}}$,$\gamma$)t}
\newcommand{\ecm}{${e}\cdot\text{cm}$}

\usepackage{rotating}
\usepackage{dcolumn}
\usepackage{bm}% bold math
\usepackage{array}
\usepackage{amsmath}
\usepackage{array}
\begin{document}

\begin{titlepage}
\vspace*{\fill}
{\centering\sffamily\Huge
\textbf{Experiments in Fundamental\\ Neutron Physics}\\
\vspace{1.0in}
\Large
Jeffrey S. Nico\\
{\it National Institute of Standards and Technology\\
Physics Laboratory\\
Gaithersburg, MD 20899\\}
\vspace{0.5in}
W. Michael Snow\\ 
{\it Indiana University and Indiana University Cyclotron Facility\\
Department of Physics\\
Bloomington, IN 47408\\}
\vspace{1.0in}
\large
December 20, 2006\par}
\vspace*{\fill}
\end{titlepage}

\pagenumbering{roman}
\tableofcontents
\clearpage
\pagenumbering{arabic}

\section{INTRODUCTION}\label{sec:Intro}
\subsection{Overview}\label{sec:Overview}

The field of neutron physics has become an integral part of investigations into an array of important issues that span fields as diverse as nuclear and particle physics, fundamental symmetries, astrophysics and cosmology, fundamental constants, gravitation, and the interpretation of quantum mechanics. The experiments employ a diversity of measurement strategies and techniques - from condensed matter and low temperature physics, optics, and atomic physics as well as nuclear and particle physics - and  address  a wide range of issues. Nevertheless, the field possesses a certain coherence that derives from the unique properties of the neutron as an electrically neutral, strongly interacting, long-lived unstable particle that can be used either as the probe or as an object of study. This review covers some of the important new contributions that neutrons have made in these diverse areas of science. By ``fundamental'' neutron physics we mean that class of experiments using slow neutrons that primarily address scientific issues associated with the Standard Model (SM) of the strong, weak, electromagnetic, and gravitational interactions and their connection with issues in astrophysics and cosmology.

Neutrons experience all known forces  in strengths that make them accessible to experimentation. It is an amusing fact that the magnitude of the average neutron interaction energy in matter, in approximately 1\,T  magnetic fields, and over a one meter fall near the surface of the Earth is the same order of magnitude ($\approx 100$\,neV). This coincidence leads to interesting and occasionally bizarre strategies for experiments that search for gravitational effects on an elementary particle.  The experiments include measurement of neutron-decay parameters,  the use of parity violation to isolate the weak interaction between nucleons, and searches for a source of time reversal violation beyond the SM. These experiments provide information that is complementary to that available from existing accelerator-based nuclear physics facilities and high-energy accelerators. Neutron physics measurements also address questions in astrophysics and cosmology. The theory of Big Bang Nucleosynthesis needs the neutron lifetime and the vector and axial vector weak couplings as input, and neutron cross sections on  nuclei are necessary for a quantitative understanding of element creation in the universe.

Free neutrons are unstable with a 15 minute lifetime but are prevented from decaying while bound in nuclei through the combined effects of energy conservation and Fermi statistics.  They  must be liberated from nuclei using nuclear reactions with MeV-scale energies to be used and studied. We define ``slow'' neutrons to be neutrons whose energy has been lowered well below this scale. The available dynamic range of neutron  energies for use in laboratory research is quite remarkable, as shown in Table~\ref{tab:terms}. Thermodynamic language is used to describe different regimes; a neutron in thermal equilibrium at 300\,K has a kinetic energy of only 25\,meV. Since its de Broglie wavelength (0.18\,nm) is comparable to interatomic distances, this energy also represents the boundary below which coherent interactions of neutrons with matter become particularly important. The most intense sources of  neutrons for experiments at thermal energies  are nuclear reactors, while accelerators can also produce higher energy neutrons by spallation.

Neutron decay is an important process for the investigation of the Standard Model of electroweak interactions. As the prototypical beta decay, it is sensitive to certain SM extensions in the charged-current electroweak sector.  Neutron decay can determine the Cabibbo-Kobayashi-Maskawa (CKM) matrix element $|V_\text{ud}|$   through increasingly precise measurements of the neutron lifetime and the decay correlation coefficients.

Searches for violations of time-reversal symmetry and/or $CP$ symmetry address issues which lie at the heart of cosmology and particle physics. Among the important issues that can be addressed by neutron experiments is the question of what mechanisms might have led to the observed baryon asymmetry of the universe. Big Bang cosmology and the observed baryon asymmetry of the universe appear to require significantly more $T$-violation among quarks in the first generation than is predicted by the Standard Model. The next generation of neutron electric dipole moment (EDM) searches, which plan to achieve sensitivities of $10^{-27}$\,\ecm\ to $10^{-28}$\,\ecm, is the most important of a class of experiments aiming to search for new physics in the $T$-violating sector.

\begin{table}
\begin{center}
\begin{tabular}{lccccc}
\hline
Term& Energy & Velocity & Wavelength & Temperature\\
 & 	 & (m/s) & (nm) & (K) &  \\
\hline
ultracold		&$<0.2$\,$\mu$eV&	$<6$	&	$>64$	&$<0.002$\\ 

very cold		& $0.2\,{\text {$\mu$eV}}\leq E < 50\,\text {$\mu$eV}$ &$6\leq v <100$& $4< \lambda\leq 64$	&$0.002\leq T<0.6$\\ 

cold	&$0.05\,{\text {meV}}<E\leq25\,\text {meV}$ &$100< v\leq2200$ &$0.18\leq \lambda< 4 $&	$0.6<T\leq300$ \\ 

thermal		&	25\,meV		&	2200		&	0.18	&	300	\\ 

epithermal	&$25\,{\text {meV}}<E\leq500\,\text {keV}$&$2200< v\leq1\times10^7$	&&\\ 

fast			&	$>500$\,keV	&	$>1\times10^7$&	&		\\ 
\hline\end{tabular}
\caption{Common terminology and spectrum of neutron energies.}
\label{tab:terms}
\end{center}
\end{table}

The last decade has also seen qualitative advances in both the quantitative understanding of nuclei, especially few body systems, and in the connection between nuclear physics and quantum chromodynamics (QCD). Low energy properties of nucleons and nuclei, such as weak interactions in n-A systems, low energy n-A scattering amplitudes, and the internal electromagnetic structure of the neutron (its electric polarizability and charge radius) are becoming calculable from the SM despite the strongly interacting nature of these systems. These theoretical developments are motivating renewed experimental activity to measure  undetermined low energy properties such as the weak interaction amplitudes between nucleons and to improve the precision of other low energy neutron measurements. The ultimate goal is to illuminate the strongly interacting ground state of QCD, the most poorly-understood sector of the SM, and to clarify the connection between the strong interaction at the hadron level and the physics of nuclei.  

This review presents and discusses the status of the experimental efforts to confront these physics questions using slow neutrons. The improvements in precision required to address these questions are technically feasible and have spurred both new experimental efforts and the development of new neutron sources. We also discuss some of the new proposed facilities under construction. It is not possible to cover the large volume of work in a review of this scope, so we refer the reader to a number of more specialized reviews wherever appropriate. Instead, we emphasize recent experiments and those planned for the near future. Neutron experiments form part of a larger subclass of low energy precision measurements which test the SM~\cite{Schreckenbach88,Dubbers99,Erler05, Nico05c,Severijns06}. There are texts that cover a broader survey of  topics  and provide historical context~\cite{Turchin65, Krupchitsky87,Alexandrov92,Byrne94}. Tests of quantum mechanics using neutron interferometry are not discussed here but are covered in detail in a recent comprehensive text~\cite{Rauch00}. Proceedings of recent international conferences also describe many planned projects or measurements in progress~\cite{Zimmer00a,Arif05}.

\subsection{Neutron Optics}

A basic introduction to neutron optics is very useful for understanding how experiments using cold and ultracold neutrons are conducted. We will use it as a unifying concept in this review, since a large number of measurements possess a transparent neutron optical interpretation. For a rigorous development of neutron optics theory, see Sears~\cite{Sears89}. 

An optical description for particle motion in a medium is possible for that subset of amplitudes in which the state of the medium is unchanged.  For these events the wave function of the particle is determined by a one-body Schrodinger equation
\begin{equation}
    \label{eq:coherent}
    \left[\frac{-\hbar^{2}}{ 2m}\Delta +v(r)\right]\psi(r)=E\psi(r),
\end{equation}
where $\psi(r)$ is the coherent wave and $v(r)$ is the optical potential of the medium. The coherent wave and the optical potential satisfy the usual coupled system of equations (Lippmann-Schwinger) of nonrelativistic scattering theory. To an excellent approximation the optical potential for slow neutrons is local and energy-independent and given by
\begin{equation}
    \label{eq:opticalpotential}
    v_{opt}(r)~=~(2\pi \hbar^2/m)\sum_{l} N_{l} b_{l},
\end{equation}
where $N_{l}$ is the number density of scatterers and $b_{l}$ is the coherent scattering length for element $l$. All neutron-matter interactions contribute to the scattering length: the dominant contributions come from the neutron-nucleus strong interaction and the interaction of the neutron magnetic moment with magnetic fields. Both interactions are spin-dependent and therefore the optical potential possesses two channels with $J=I\pm 1/2$. The coherent scattering length $b$ is the sum of the scattering lengths in both scattering channels weighted by the number of spin states in each channel.  From a quantum mechanical point of view, this is the total amplitude for a neutron to scatter without a change in the internal state of the target. The effect of the optical potential for a nonabsorbing uniform medium with positive coherent scattering length is to slow down the neutrons as they encounter the potential step due to the matter, thereby decreasing the neutron wave vector, \textit{K}, within the medium. A neutron index of refraction can be defined by this relative change in the magnitude of the wavevector $n=K/k$. Conservation of energy at the boundary determines the relation to the optical potential
\begin{equation}
    \label{eq:refractionindex}
    n^{2}=1-{v_{0}/E}.
\end{equation}
Incoherent effects due to neutron absorption and incoherent scattering with the medium remove probability density from the coherent wave and can be modeled by adding an imaginary part to the optical potential whose magnitude is determined by the optical theorem of nonrelativistic scattering theory. The scattering lengths therefore become complex.

It may seem counterintuitive that the attractive neutron-nucleus interaction can give rise to a repulsive neutron optical potential. For a weak potential there is indeed a one-to-one relation between the sign of the zero-energy scattering phase shift and the sign of the potential, but the neutron-nucleus interaction is strong enough to form bound states. If one recalls Levinson's theorem from nonrelativistic scattering theory, which says that the zero energy phase shift increases by $\pi$ for each bound state, one can see that it is possible for an attractive potential to produce a negative phase shift. Simple models~\cite{Peshkin71} and more detailed treatments~\cite{Aleksejev98} of the neutron-nucleus interaction show that for most nuclei the scattering lengths are indeed positive. A summary of all neutron scattering lengths up to 1991~\cite{Koester91} and a set of recommended values~\cite{Rauch00b} exist in the literature.

The fact that the neutron optical potential is repulsive for most materials has a number of far-reaching consequences for slow neutron technology that are briefly mentioned here. Since the neutron index of refraction is less than unity, the well-known light optical phenomenon of total {\it internal} reflection for rays at an interface becomes total {\it external} reflection for neutrons. Neutron guides~\cite{Maier63}, the neutron analog of optical fibers, are used to conduct neutron beams far from the neutron source by total external reflection from mirrors with negligible loss in intensity within their phase-space acceptance. The phase-space acceptance can be increased over that of a uniform medium by multilayer coatings called supermirrors, which expand the effective critical angle for total external reflection through constructive interference of scattering from different layers.  

If the mirror is made of magnetic material with a magnetic scattering amplitude of the same size as the nuclear scattering amplitude, the optical potential can be large for one neutron spin state and close to zero for the opposite spin state. In this case the reflected neutron beam is highly polarized. Supermirror neutron polarizers which polarize essentially all of the phase space of a cold neutron beam transported by a neutron guide to greater than 98\,\% are routinely available. A polarized neutron beam can also be produced through transmission of an unpolarized neutron beam through a polarized medium; both polarized hydrogen (using spin-dependent scattering) and polarized \Het\ (using spin-dependent absorption) have been used.       

Neutrons whose kinetic energy is lower than the optical potential of the medium that are incident on a surface will be reflected at all angles of incidence. Such neutrons can therefore be trapped in a material bottle through total external reflection and are called ultracold neutrons (UCN). The possibility of performing measurements on an ensemble of free neutrons almost at rest in the lab is a decisive advantage in certain experiments. Neutrons in this energy range can also be confined by magnetic field gradients that act on the neutron magnetic moment. In addition, such neutrons can only rise a meter or so in the gravitational field of the Earth. In both of these cases the ``optical potential'' is simply the normal potential energy of the neutron in the presence of the relevant external field.   

The expression for the low energy neutron scattering length of an atom away from nuclear resonances for unpolarized atoms and neutrons and nonmagnetic materials is
\begin{equation}
    \label{eq:alllengths}
    b=b_{nuc}+Z(b_{ne}+b_{s})[1-f(q)]+b_{pol},
\end{equation}
where $b_{nuc}$ is the scattering amplitude due to the neutron-nucleus strong force, $b_{ne}$ is the neutron-electron scattering amplitude due to the internal charge distribution of the neutron, $b_{s}$ is the Mott-Schwinger scattering due to the interaction of the magnetic moment of the neutron with the ${\bf
v} \times {\bf E}$ magnetic  field seen in the neutron rest frame from electric fields, $b_{pol}$  is the scattering amplitude due to the electric polarizability of the neutron in the intense electric field of the nucleus, and $f(q)$ is the charge form factor (the Fourier transform of the electric charge distribution of the atom). The electromagnetic contribution to the scattering lengths from $b_{ne}$ and $b_{s}$ are exactly zero for forward scattering due to the neutrality of the atoms, which forces the charge form factor $f(q) \to 1$ as $q \to 0$~\cite{Sears89}. The weak interaction also makes a contribution to the scattering length which possess a parity-odd component proportional to ${\vec{s_{n}} \cdot {\vec{p}}}$, where ${\vec{s_{n}}}$ is the neutron spin and ${\vec{p}}$ is the neutron momentum. The relative sizes of these contributions to the scattering length from the strong, electromagnetic, and weak interactions are roughly $1:10^{-3}:10^{-7}$.

\subsection{Neutron Sources}\label{sec:Sources}

Most fundamental neutron physics experiments are conducted with slow neutrons for two main reasons. First, slower neutrons spend more time in an apparatus. Second, slower neutrons can be more effectively manipulated through coherent interactions with matter and external fields. Free neutrons are usually created through either fission reactions in a nuclear reactor or through spallation in high $Z$ targets struck by GeV proton beams. We briefly examine these neutron sources and the process by which cold and ultracold neutrons (UCN) are produced starting from neutrons with energies several orders of magnitude greater.

Neutrons are produced from fission in a research reactor at an average energy of  2\,MeV. They are slowed to thermal energy in a moderator (such as heavy or light water, graphite, or beryllium) surrounding the fuel. The peak core fluence rate of research reactors is typically in the
range $10^{14}$\ncms\ to $10^{15}$\ncms.  To maximize the neutron density, it is necessary to increase the fission rate per unit volume, but the power density is ultimately limited by heat transfer and material properties. In the spallation process, protons (typically) are accelerated to  energies in the  GeV range and strike a high $Z$ target, producing approximately 20 neutrons per proton with energies in the fast and epithermal region~\cite{Windsor81}. This is an order of magnitude more neutrons per nuclear reaction than from fission. Existing spallation sources yield peak neutron rates of $10^{16}$\,s$^{-1}$ and $10^{17}$\,s$^{-1}$. Although the time-averaged fluence from spallation neutron sources is presently about an order of magnitude lower than for fission reactors, there is potentially more room for technical improvements in the near-term future due in part to the relaxation of the constraints needed to maintain a nuclear chain reaction~\cite{Difilippo91} and to the potential for future developments in GeV proton accelerator technology.

The main feature that differentiates spallation sources from reactors is their convenient operation in a pulsed mode. At most reactors one obtains continuous beams with a thermalized Maxwellian energy spectrum. In a pulsed spallation source, neutrons arrive at the experiment while the production source is off, and the frequency of the pulsed source can be chosen so that slow neutron energies can be determined by time-of-flight methods.  The lower radiation background and convenient neutron energy information can be advantageous for certain experiments. The frequency is typically chosen to lie in the 10\,Hz to 60\,Hz range, so that the subsequent neutron pulses from the moderator do not overlap for typical neutron spectra and distances to the experimental area.

Fast neutrons reach the thermal regime most efficiently through a logarithmic energy cascade of roughly 20 to 30 collisions with matter rich in hydrogen or deuterium. Neutrons are further cooled by a cryogenic neutron moderator adjacent to the reactor core or spallation target held at a temperature of $\approx 20$\,K.  One generally wants the moderator as cold as possible to increase the phase space density of the neutrons.   As the neutron wavelengths become large compared to the atomic spacings, the total scattering cross sections in matter are dominated by elastic or quasielastic processes, and it becomes more difficult for the neutrons to thermalize. The development of new types of cold neutron moderators is therefore an important area for research. 

It is not practical to describe specific neutron facilities in any detail in this review, but we note a few where the bulk of research efforts have been carried out. For thirty years the most active facility for fundamental neutron research has been the Institut Max von Laue - Paul Langevin (ILL) in Grenoble, France~\cite{ILL01}. Its 58\,MW reactor is the focal point of neutron beta-decay and UCN physics in the world, and its cold neutron beam H113 for fundamental physics, the first large ballistic supermirror neutron guide, possesses the highest cold neutron intensity in the world~\cite{Hase02, Abele06}. The new FRM-II reactor has come online in Munich, and the experimental program in fundamental neutron physics has commenced~\cite{Zimmer05a}. The most active institutions in the US are the National Institute of Standards and Technology (NIST)~\cite{Nico05a}, which operates the NG-6 end position, two monochromatic neutrons beams at 0.49\,nm and 0.9\,nm, and a neutron interferometer facility on the NG-7 guide,  and Los Alamos National Laboratory (LANL), which operates the FP12 beamline~\cite{Seo04}. In Russia, there are significant efforts at the Petersburg Nuclear Physics Institute (PNPI), Joint Institute for Nuclear Research (JINR) in Dubna, and the Kurchatov Institute and in Japan at KEK.  Many smaller sources play an essential role in the development of experimental ideas and techniques. 

\begin{table}
\begin{center}
\caption{Some operating parameters for major cold neutron reactor-based user facilities with active fundamental physics programs. Approximate fluence rates are given as neutron capture fluence.}
\label{tab:sources}
\begin{tabular}{lcccc}
\hline
Parameter& ILL&ILL& NIST & FRM-II \\
 & PF1 & PF2 & NG-6 & Mephisto  \\
\hline
Power (MW)	&58& 58& 20& 20 \\ 

Guide length (m)	&60& 74&68& 30 \\

Guide radius (m)	&4000& 4000& $\infty$& 460 \\

Guide type ($m=$)	&1.2& 2& 1.2 & 3 \\

Cross section (cm$^2$)	&$6\times 12$& $6\times 20$& $6\times 15$& $5\times 11.6$\\

Fluence rate ($\times 10^9$\,\ncms)	&4 &14 &2 & 7 \\

\hline
\end{tabular}
\end{center}
\end{table}

In addition to the existing sources, the last decade has seen tremendous growth in the construction of facilities and beamlines devoted to fundamental neutron physics. Many of these new facilities are at spallation sources.  The Paul Scherrer Institut (PSI) operates the Swiss Spallation Neutron Source SINQ~\cite{Fischer97, Bauer98}  as a continuous spallation source and has constructed a cold neutron beamline FUNSPIN dedicated to fundamental physics~\cite{Schebetov03, Zejma05}.  In the US, the 2\,MW Spallation Neutron Source (SNS) is under construction, and the fundamental physics beamline (FNPB) should be operational some time in 2008~\cite{Greene05}.  The Japanese Spallation Neutron Source (JSNS) is in the construction phase and is also anticipated to become operational in 2008. A fundamental neutron physics beamline at the JSNS with a direct line-of-sight to the neutron moderator would be complementary to the (curved) SNS beam and would allow for experiments using pulsed epithermal neutrons. Tables~\ref{tab:sources} and \ref{tab:pulsedsources}  give a few of the measured (or projected) cold-neutron beam properties for some of the facilities with active fundamental physics programs. As a word of caution, it is not always straightforward to compare the cold neutron beam properties of different sources, and experimentalists must consider all the parameters before determining the optimal source. 

In the neutron energy spectrum from a cold moderator there is a very small fraction whose energies lie below the $\approx 100$\,neV neutron optical potential of matter. Such neutrons are called ultracold neutrons, and they can be trapped by total external reflection from material media. The
existence of such neutrons was established experimentally in the late 1960s~\cite{Luschikov69,Steyerl69}. The UCN facility at the ILL employs a turbine to mechanically convert higher energy neutrons to UCN~\cite{Steyerl86}. Although the density of neutrons is bounded by the original phase space in the source (Liouville's theorem), this technique produces enough UCN to conduct a number of unique and fundamental experiments described in part below. Over the last decade new types of UCN converters have been developed that can increase the phase space density through the use of ``superthermal''  techniques~\cite{Golub77}.  They involve energy dissipation in the moderating medium (through phonon or magnon creation) and can therefore increase the phase space density of the UCN since  Liouville's theorem does not apply.  Superfluid helium~\cite{Ageron78} and solid deuterium~\cite{Saunders04}  have been used most successfully as superthermal UCN sources, and other possible UCN moderating media such as solid oxygen are also being studied~\cite{Liu06}. The lack of neutron absorption in \Hef\ along with its other unique properties makes possible experiments in which the measurement is conducted within the moderating medium. These developments have led to new construction projects for UCN facilities at LANL~\cite{Saunders04}, PSI~\cite{Atchison05}, FRM~\cite{Trinks00}, KEK~\cite{Masuda02}, NC State~\cite{Korobkina02}, Mainz, and other sources.  Extensive treatments of UCN physics are found in References~\cite{Ignatovich90,Golub91}.

\begin{table}
\begin{center}
\caption{Some operating parameters for major cold neutron spallation-source user facilities with active (and proposed) fundamental physics programs. Approximate time-averaged fluence rates are given as neutron capture fluence. Note that for some experiments the peak fluence rather than the time-averaged fluence is the more important parameter.}
\label{tab:pulsedsources}
\begin{tabular}{lcccc}
\hline
Parameter& SINQ&LANSCE&(SNS)&(JSNS)\\
 & FunSpin	 & FP12 & (FNPB) &\\
\hline
Time-averaged current (mA)	& 1.2& 0.1&(1.4) & (0.3)\\

Source rep. rate (Hz)	& dc &20 & (60)& (25) \\

Guide length (m)	&7&8 & (15)& (10 to 20)\\

Guide radius (m)	&$\infty$ & $\infty$& $(117)$&($\infty$)\\

Guide type ($m=$)	& 3&3&(3.5) &(3)\\

Cross section (cm$^2$)	&$4\times 15$	& $9.5\times 9.5$& ($10\times 12$)&($10\times 10$)\\

Fluence rate ($\times 10^8$\,\ncms)	& 8 & & (10) & (5) \\

\hline
\end{tabular}
\end{center}
\end{table}

\section{NEUTRON DECAY AND STANDARD MODEL TESTS}\label{sec:SMThy}

There are a several reviews that discuss weak interaction physics using slow neutrons in greater detail or provide additional information~\cite{Byrne82,Dubbers91a,Pendlebury93}.  A recent publication emphasizes the issue of CKM unitarity~\cite{Abele03}.  A comprehensive review of measurements in neutron and nuclear beta decay to test the Standard Model in the semileptonic sector and its possible extensions along with a comparison with other probes of similar physics has appeared~\cite{Severijns06}.
 
\subsection{Theoretical Framework}\label{sec:NDBThy}

The neutron is composed of two down quarks and an up quark, and it is stable under the strong and electromagnetic interactions, which conserve quark flavor. The weak interaction can convert a down quark into an up quark through the emission of the $W$ gauge boson. The mass difference of the neutron and proton is so small that the only possible decay products of the $W$ are an electron and antineutrino with the release of energy distributed among all the decay products: $n\rightarrow p+e{^-}+\bar{\nu}_e+0.783\,\text {MeV}$. There are two other decay channels available to the neutron. There is the radiative decay mode with a photon in the final state and decay to a hydrogen atom and an antineutrino~\cite{Nemenov80a, Nemenov80b, Byrne01}. Very recently, the radiative decay mode has been definitively observed and the branching ratio measured at the 10\,\% level~\cite{Nico06}. The decay into a hydrogen atom has not yet been seen although a search for $n \to H+\bar{\nu}_e$ is under investigation~\cite{Schott06}. Beta-decay experiments can test the assumptions of the SM by performing precision measurements on the decay product energies,  momenta, and their correlations with the neutron spin.

Free neutron decay in the Standard Model is described in lowest order by a mixed vector/axial-vector current characterized by two coupling strengths, \gv\ and \ga, the vector and axial-vector
coupling coefficients.  Since the momentum transfers involved in neutron beta decay are small compared to the $W$ and $Z$ masses, one can write an effective Lagrangian that describes neutron decay in the Standard Model as a four-fermion interaction 
\begin{equation}
\label{eq:lagrangian}
\mathcal{L}_{int}={G_{F}V_\text{ud} \over 2\sqrt{2}}(V_{\mu}-\lambda A_{\mu})(v^{\mu}-a^{\mu}),
\end{equation}
where $V_{\mu}=\overline{\psi_{p}}\gamma_{\mu}\psi_{n}$, $v^{\mu}=\overline{\psi_{e}}\gamma^{\mu}\psi_{\nu}$, 
$A_{\mu}=\overline{\psi_{p}}\gamma_{\mu} \gamma_{5}\psi_{n}$, and 
$a^{\mu}=\overline{\psi_{e}}\gamma^{\mu} \gamma_{5}\psi_{\nu}$
are the hadronic and leptonic vector and axial vector currents constructed from the neutron, proton, electron, and neutrino fermion fields, $G_{F}$ is the Fermi decay constant, $V_\text{ud}$ is a CKM matrix element, and $\lambda$ is the ratio of the axial vector and vector couplings. 

The V-A structure for the weak currents is incorporated directly into the standard electroweak theory by restricting the weak interaction to operate only on the left-handed components of the quark and lepton fields. A more fundamental understanding of the reason for this parity-odd structure of the weak interaction is still lacking. There is also no physical understanding for the measured values of the CKM mixing matrix elements between the quark mass eigenstates and their weak interaction eigenstates. The fact that the matrix is unitary is ultimately a consequence of the universality of the weak interaction gauge theory. Extensions to the SM which either introduce non V-A weak currents or generate violations of universality can therefore be tested through precision measurements in beta decay. A recent reanalysis of the constraints on non V-A charged currents showed that improved neutron decay measurements have set new direct limits on such couplings, which are typically constrained at the 5\,\% level~\cite{Severijns06}. Complementary constraints on non V-A charged currents in neutron beta decay from a combination of neutrino mass limits and cosmological arguments have recently appeared~\cite{Ito05}   

\begin{figure}
\begin{center}
%\epsfxsize10pc
%\centerline{\epsfbox{fig1.eps}}
\includegraphics[width=6.5in]{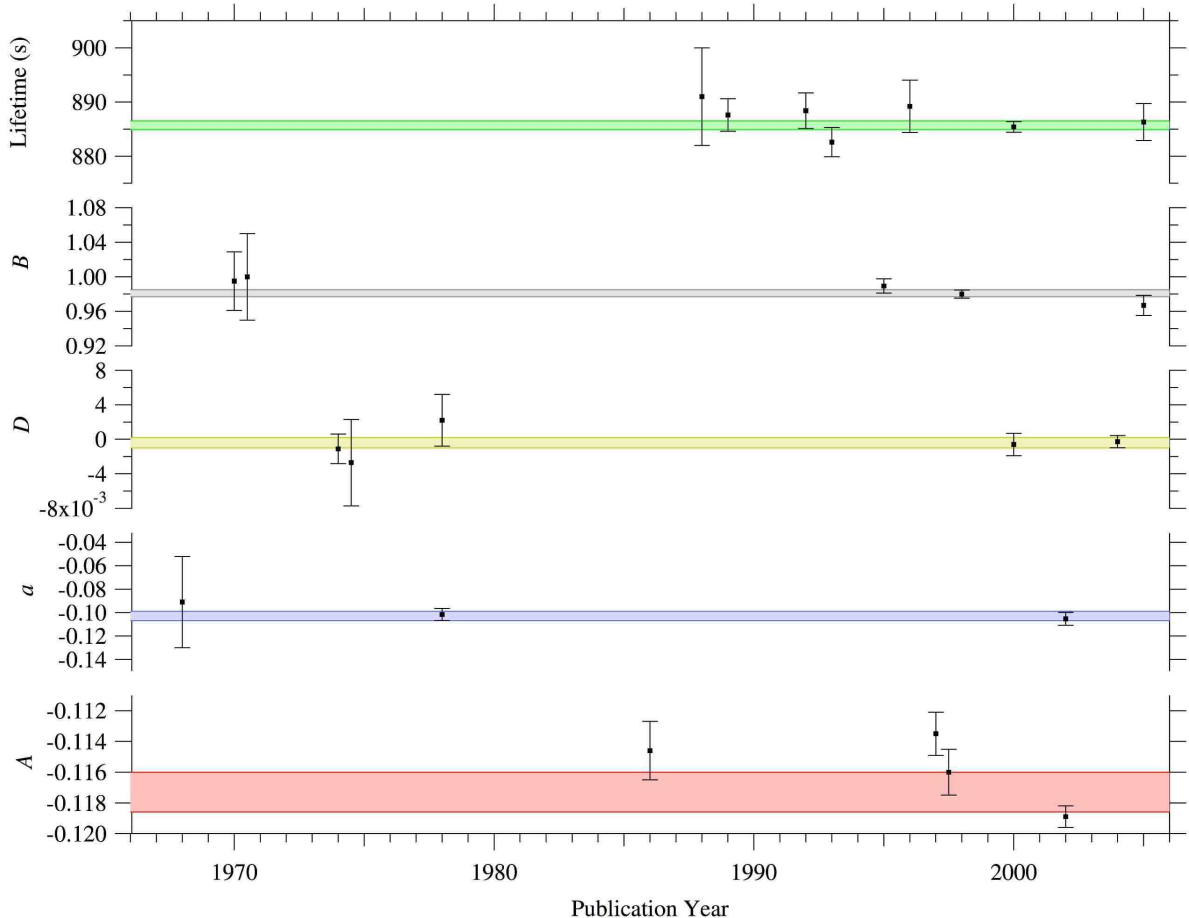}
\caption{A summary plot of the measurements of the neutron lifetime and correlation coefficients that are used in the 2006 compilation of the PDG. The shaded bands represent the $\pm 1\,\sigma$ region.}
\label{fig:tau_coeffs}
\end{center}
\end{figure}

The probability distribution for beta decay in terms of the neutron spin and the energies and momenta of the decay products~\cite{Jackson57} can be written
\begin{equation}
    \label{eq:Jackson}
        dW \propto (g_{V}^{2} + 3 g_{A}^{2}) F(E_{e}) \left[1+
          a\frac{\vec{p}_{e}\cdot\vec{p}_{\nu}} {E_{e}E_{\nu}}+ b\frac{{m}_{e}} {E_{e}}
          + {\vec{\sigma}_{n}} \cdot \left( A\frac{\vec{p}_{e}} {E_{e}}+
          B\frac{\vec{p}_{\nu}} {E_{\nu}} + D\frac{\vec{p}_{e} \times
          \vec{p}_{\nu}}{E_{e}E_{\nu}} \right)\right],
\end{equation}
where one defines
\begin{align*}
	\tau_n & = \frac{2 \pi^3 \hbar^7}{m_e^5 c^4}\frac{1}{f(1+\delta_R)({g_V}^2+3{g_A}^2)} & \quad &= (885.7 \pm  0.8)~\text{s} && \text{neutron lifetime} \\
	\lambda & = \left|\frac{g_A}{g_V}\right|e^{i \phi} & \quad & =  -1.2695 \pm 0.0029 && \text{coupling constant  ratio} \\
	a & = \frac{1-|\lambda|^{2}}{1+3|\lambda|^{2}} & \quad & = -0.103 \pm 0.004 && \text{electron-antineutrino asymmetry} \\
 	b & = 0 & \quad  & =0 && \text{Fierz interference}\\
	A & = -2 \frac{|\lambda|^{2}+|\lambda|\cos\phi}{1+3|\lambda|^{2}} & \quad &  = -0.1173 \pm 0.0013 && \text{spin-electron asymmetry} \\
	B & = 2 \frac{|\lambda|^{2}-|\lambda|\cos \phi}{1+3|\lambda|^{2}} & \quad & =0.981 \pm 0.004 &&
\text{spin-antineutrino asymmetry} \\
	D & = 2 \frac{|\lambda|\sin \phi}{1+3|\lambda|^{2}} & \quad & =(-4 \pm 6) \times 10^{-4} && \text{$T$-odd triple-product.}
\end{align*}
In these equations, which neglect small contributions  such as weak magnetism and radiative corrections, $F(E_e)$ is the electron energy spectrum, $\vec{p}_e$, $\vec{p}_{\nu}$, $E_e$, and $E_{\nu}$ are the momenta and kinetic energies of the decay electron and antineutrino,
${\vec{\sigma}_n}$ is the initial spin of the decaying neutron, and $\phi$ is the phase angle between the weak coupling constants \ga\ and \gv. $f$ is the statistical rate function and $\delta_R$ is a radiative correction evaluated for the neutron; the product $f(1+\delta_R)=1.71489\pm0.00002$ has been calculated with high precision~\cite{Towner95}.  The spin-proton asymmetry correlation coefficient $C$ is proportional to the quantity $A+B$. The values represent the world averages as compiled by the Particle Data Group (PDG)~\cite{Yao06}.  The parameter $\lambda$ can be extracted from measurement of either $a$, $A$, or $B$.  If the neutron lifetime $\tau_{n}$ is also measured, \gv\ and \ga\ can be determined uniquely under the assumption that $D=0$. Radiative corrections to the correlation coefficients exist and have been evaluated recently using  effective field theory (EFT) based methods~\cite{Ando04, Torres04, Bernard04a, Bernard04b,Bunatian05, Gudkov05}. Figure~\ref{fig:tau_coeffs} shows the recent history of measured values of the lifetime and correlation coefficients as used by the PDG~\cite{Yao06}.

Accurate determination of the parameters that describe neutron decay can provide important information regarding the completeness of the three-family picture of the SM through a test of the unitarity of the CKM matrix, or $|V_{\mathrm{ud}}|^2 +|V_{\mathrm{us}}|^2 + |V_{\mathrm{ub}}|^2 = 1$. One strong motivation for more accurate measurements of neutron decay parameters is to measure $|V_{\mathrm{ud}}|$, the largest term in the unitarity sum. The matrix elements $|V_{\mathrm{us}}|$ and $|V_{\mathrm{ub}}|$ are obtained from high energy accelerator experiments.

The conserved-vector-current hypothesis implies that $g_{\mathrm{V}}={G_{\mathrm{F}}|V_{\mathrm{ud}}|(1+\Delta^{\rm{V}}_{\rm{R}})}$, where $\Delta^{\rm{V}}_{\rm{R}}$ is the nucleus-independent radiative correction. One can determine $|V_{\mathrm{ud}}|$ by measuring the $\mathcal{F}t$ values  of superallowed $0^{+} \to 0^{+}$ $\beta$ transitions between isobaric analog states~\cite{Hardy05}.
Recently, a new method for calculating hadronic corrections to electroweak radiative corrections that combines high-order perturbative QCD calculations with a large N-based extrapolation has been applied to superallowed nuclear beta decay~\cite{Marciano06}. This calculation reduced the theoretical loop uncertainty by approximately a factor of two and now yields a value of $| V_{\mathrm{ud}} | = 0.97377 \pm 0.00027$ with the uncertainty dominated by theoretical nucleus-dependent radiative corrections. Pion beta decay is theoretically the cleanest system in which to measure $|V_{\mathrm{ud}}|$, but the small branching ratio has so far precluded a measurement with enough sensitivity to compete with superallowed beta decay and neutron decay. The latest measurement from pion beta decay gives $|V_{\mathrm{ud}}|=0.9728 \pm 0.0030$~\cite{Pocanic04}.

Using values of $|V_{\mathrm{us}}|$ and $| V_{\mathrm{ub}} |$ taken from the current recommendations of the PDG,  one obtains $|V_{\mathrm{ud}}|^2 + |V_{\mathrm{us}}|^2 + |V_{\mathrm{ub}}|^2 = 0.99919 \pm 0.00108$, which removes the more than two standard deviation discrepancy with unitarity~\cite{Towner03}. The primary reason for the significant shift is a new value of $|V_{\mathrm{us}}|$ that is about 2.5\,\% larger than the previous PDG evaluation. The result is based on recent experiments in semileptonic kaon decays, including measurements of the branching ratios, form factors, and lifetime~\cite{Franzini04}. (There are also renewed theoretical investigations to extract
$| V_{\mathrm{us}}|$ from hyperon decay~\cite{Cabibbo04} and precision experiments in hadronic tau decays~\cite{Gamiz05}. ) It still uses the value of $f_{+}(0)$ calculated by Leutwyler and Roos~\cite{Leutwyler84}, but one should note that other calculations exist that produce numbers that differ by as much as 2\,\%. The value of $|V_{\mathrm{ub}}|$ has also changed, but it has a negligible contribution to the unitarity sum.

\begin{figure}[t]
\begin{center}
%\epsfxsize10pc
%\centerline{\epsfbox{fig2.eps}}
\includegraphics[width=5.4in]{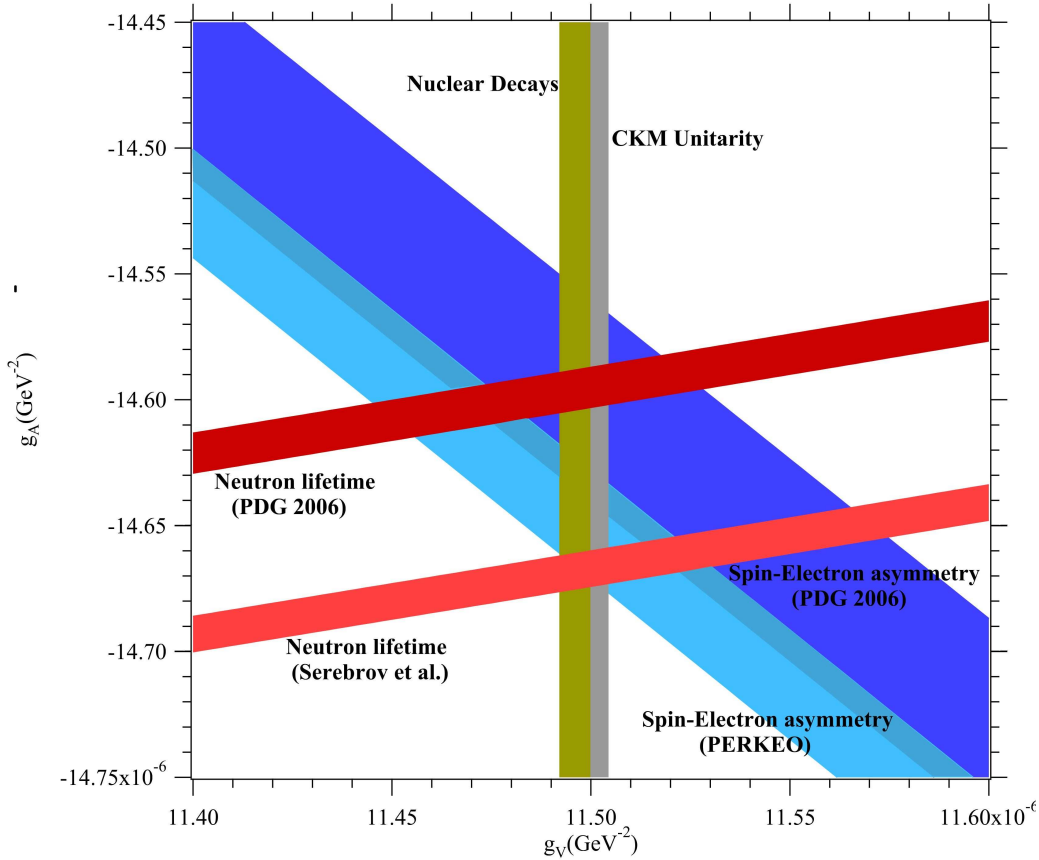}
\caption{The weak coupling constants \gv\ and \ga\  determined from neutron decay parameters, superallowed $0^+ \rightarrow 0^+$  nuclear decay transitions, and CKM unitarity. The bands represent $\pm 1$ standard deviation uncertainties.   For the neutron decay parameters, the uncertainties are dominated by experimental uncertainties. For the superallowed $0^+ \rightarrow 0^+$ nuclear decays, the uncertainty is dominated by radiative correction calculations. In addition to the recommended values of the PDG 2006 evaluation shown for neutron decay, the PERKEO result~\cite{Abele02} and the lifetime measurement of Serebrov {\it et al}.~\cite{Serebrov05a} are shown independently for comparison.}
\label{fig:unitarity}
\end{center}
\end{figure}

Neutron beta decay offers a somewhat cleaner theoretical environment for extracting \gv\ than the superallowed $0^+ \rightarrow 0^+$  $\beta$ transitions due to the absence of other nucleons (although some radiative corrections are common to both systems).  In neutron decay the value of $|V_{\mathrm{ud}}|$ is obtained from the expression
\begin{equation}
|V_{\mathrm{ud}}|^2= {1\over f(1+\delta_{\rm{R}})\tau_n}{K/{\mathrm {ln}} 2\over G_{\rm{F}}(1+\Delta^{\rm{V}}_{\rm{R}})(1+3\lambda^2)},
\end{equation}
where $K$ is a constant. Marciano and Sirlin recently refined calculations of the radiative corrections~\cite{Marciano06} to yield
\begin{equation}
|V_{\mathrm{ud}}|^2= {(4908.7 \pm 1.9)\,s\over \tau_n (1+3\lambda^2)}.
\end{equation}
Thus, with the lifetime and the parameter $\lambda$, most precisely obtained from measurements of the neutron spin-beta asymmetry angular correlation, $|V_{\mathrm{ud}}|$ can be determined uniquely. Using the PDG values of $\tau_{n}$ and $\lambda$, the same unitarity test gives $\sum_i |V_{\mathrm{ui}}|^2 = 0.9971 \pm 0.0039$,  consistent with unity but less precise.  The present situation regarding unitarity is summarized in Figure~\ref{fig:unitarity}. Although the values from the PDG evaluation imply good agreement among the neutron measurements, the nuclear systems, and unitarity, the plot also shows unresolved disagreements in the neutron lifetime value and the spin-electron asymmetry. Until those issues are resolved, the results from the neutron system remain in question.

A precision determination of $|V_{\mathrm{ud}}|$ should be seen in the context of the overall effort  to determine with high precision all the parameters of the CKM matrix. At CLEO-c it should be possible to measure the CKM matrix element $|V_{\mathrm{cd}}|$ to 1\,\% accuracy if lattice gauge theory calculations of the required form factors can match the expected precision of the data~\cite{Shipsey02,Shipsey03}. Recent developments in lattice gauge theory have produced a qualitative improvement in the accuracy with which certain observables can be calculated; for example, the latest result for $V_{us}=0.2219\pm 0.0026$ from the MILC collaboration rivals the experimental precision~\cite{Aubin04}.  Successful lattice gauge theory calculations of form factors for charmed quark decays would make possible another independent check of CKM unitarity using the first column, $|V_{\mathrm{ud}}|^2+|V_{\mathrm{cd}}|^2+|V_{\mathrm{td}}|^2=1$. 

\subsection{Neutron Lifetime}\label{sec:Lifetime}
\subsubsection{Recent Lifetime Experiments}\label{sec:recentlifetime}

Two distinct strategies have been employed to measure the neutron lifetime. The first measurements were performed using thermal or cold beams of neutrons and measured simultaneously both the average number of neutrons $N$ and the rate of neutron decays ${dN/dt}$ from a well-defined fiducial
volume in the beam.  All modern experiments in neutron decay use low energy neutrons because more particles decay within a given volume, thus increasing the statistical power. The beam technique requires absolute knowledge of the efficiencies to extract the lifetime from ${dN/dt}=-N/{\tau_{n}}$. A significant improvement in the precision was obtained by implementing a segmented proton trap to better define the fiducial volume~\cite{Byrne89}. The two more precise experiments used such a proton trap as well as a neutron detector with an efficiency that was inversely proportional to the neutron velocity~\cite{Byrne96,Dewey03}. These experiments are currently limited by systematic effects related to neutron counting, but reductions in the uncertainties are quite feasible, and an effort is underway to improve the precision of one of the beam experiments.

With the advent of UCN production techniques, a second approach to lifetime measurements was developed that confines neutrons in material ``bottles'' or magnetic fields and measures the number of neutrons remaining as a function of time.  At a time $t$ the number of neutrons is given by $N(t)=N(0)e^{-t/\tau_{\rm{n}}}$, so by measuring $N(t)$ at different times, one can determine the neutron lifetime.  With this technique one avoids the necessity of knowing the absolute neutron density and detector efficiencies but other systematic effects arise. The measured lifetime is given by $\tau_{\rm{tot}}^{-1}=\tau_{\rm{n}}^{-1} +\tau_{\rm{loss}}^{-1}$ and includes losses from the confinement cavity as well as neutron decay. To isolate $\tau_{loss}$, which is typically dominated by nonspecular processes in the neutron interaction with the walls of the confinement vessel, one measures the lifetime in bottles with different surface-to-volume ratios and performs an extrapolation to an infinite volume (or equivalently zero collision rate).  These losses depend on the UCN energy spectrum, which can change during the storage interval, so much work has been done to understand the spectrum evolution and loss mechanisms and to find surface materials with lower loss probabilities. To address losses experimentally, one group simultaneously measured the UCN storage time and the inelastically scattered neutrons~\cite{Arzumanov00}, thus monitoring the primary loss process. Because neutrons can also occupy unbound metastable orbits in the trap, it is necessary to ``clean'' the phase space of the UCN to remove them so that the remaining losses are dominated by wall interactions.   

\begin{figure}
  \includegraphics[width=6in]{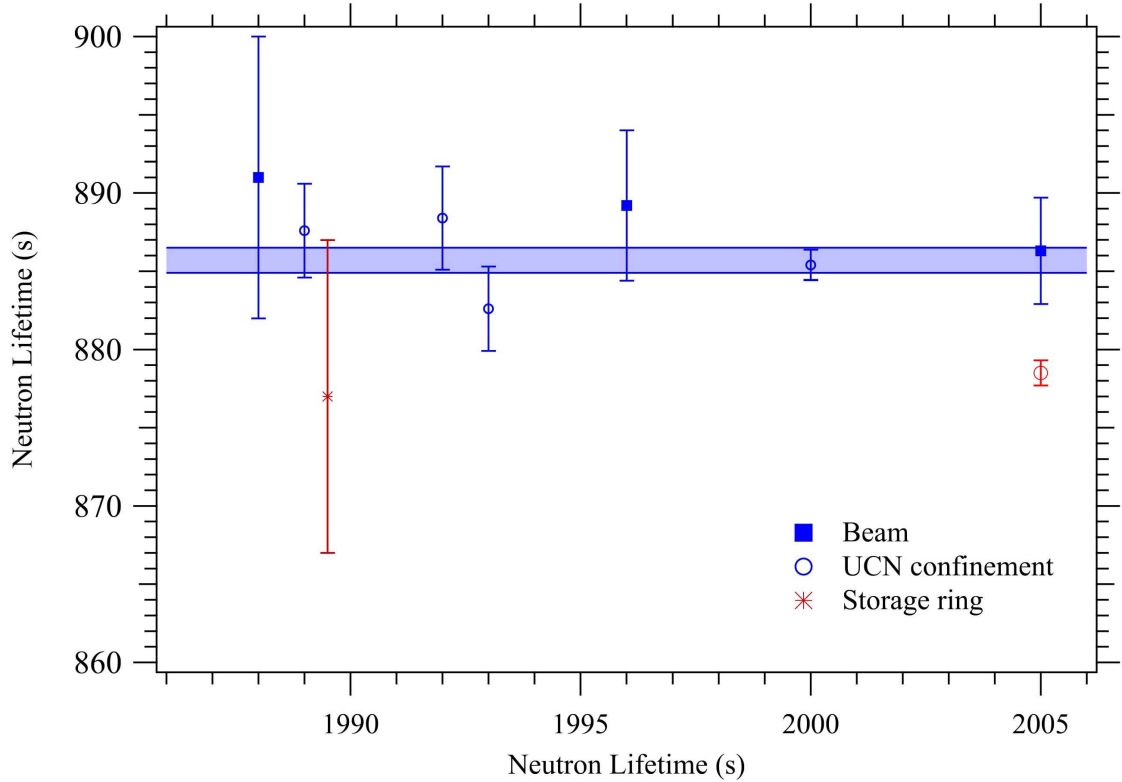}
  \caption{\label{fig:Lifetimes}A comparison of recent lifetime measurements. The shaded band represents the $\pm 1$ standard deviation for the seven measurements (shown in blue) included in the PDG evaluation.}
\end{figure}

Accurate measurements using each of these independent methods are important for establishing the reliability of the result for \taun, particularly given a recent measurement that is significantly discrepant with the world average from the PDG 2006 evaluation. Figure~\ref{fig:Lifetimes} shows a summary of measurements over the last 20 years with competitive uncertainties.  (A more historical and detailed review of neutron lifetime experiments is found in Ref.~\cite{Schreckenbach92}.) Seven of the  experiments~\cite{Spivak88,Mampe89,Nez92,Mampe93,Byrne96,Arzumanov00,Nico05b} contribute to a current neutron lifetime world average of $\tau_{n} = (885.7\pm0.8)$~s~\cite{Yao06}. The agreement among the measurements is very good. The experiments using beams do not have as much statistical influence as those using UCN, but they are also in good agreement.

Also shown on Figure~\ref{fig:Lifetimes} is the most recent experiment of Serebrov {\it et al}. that produced a value of $\tau_{n} = (878.5\pm0.8)$~s~\cite{Serebrov05a}. The result is more than six standard deviations away from the PDG average. The experiment used ultracold neutrons confined both gravitationally and by a material bottle and measured the total storage lifetime. The loss rate on the walls is given by $\tau_{\rm{loss}}^{-1} = \eta\gamma$, where $\eta$ is a wall-loss coefficient related to the probably of losing a neutron per interaction with the wall and $\gamma$ is the loss-weighted collision frequency. Other loss mechanisms were considered negligible. The neutron lifetime was  determined by extrapolating to zero collision frequency. The experiment used low temperature Fomblin oil to significantly reduce the value of $\eta$ in comparison with previous experiments. The improvement reduced the size of the extrapolation necessary to extract \taun, thus making the experiment less sensitive to some systematic effects. They also used two storage vessels with different surface-to-volume ratios and were able to change the energy distribution of the confined UCN, both of which change the collision frequency. The group intends to make additional measurements with a variable-volume trap to change the collision frequency while maintaining the same trap surface and vacuum conditions.

The substantial difference between the neutron lifetime of PDG average and that of Serebrov {\it et al}. is not understood. It is essential to resolve the disagreement, and it can only be accomplished through new measurements~\cite{Nico06a}. The issue of how the value of $|V_{\mathrm{ud}}|$ determined from neutron beta decay parameters \taun\ and $\lambda$ compares with those from the superallowed decays and CKM unitarity can not be fully addressed without a resolution of the value of \taun. 

\subsubsection{Prospects in Neutron Lifetime Measurements}\label{sec:newtau}

Several new experimental efforts to measure \taun\ are underway that include both refinement of existing experiments and techniques as well as development of new approaches. Most of the proposed experiments use confined UCN since the systematic effects are thought to be less challenging. The beam experiments are important because their set of systematic effects is almost completely orthogonal to those of the UCN confinement experiments. Agreement using the two techniques is needed for reliability of the value of the lifetime.

One experiment proposes to use a low temperature Fomblin oil to reduce the UCN collisional losses . Recent data demonstrate a UCN reflection loss coefficient of $5\times 10^{-6}$ per bounce  when the vessel temperature is in the range of 105\,K to 150\,K. They intend to use this surface coating in an ``accordion-like'' storage vessel, thus allowing one to vary the trap volume while keeping the surface area and characteristics constant~\cite{Yero05}. The collaboration expects to achieve a precision of 1\,s. Another experiment will store UCN magnetically in vacuum using an arrangement of permanent magnets and superconducting solenoids and extract the protons electrostatically. The lifetime is measured by real-time detection of the decay protons and counting the integral number of neutrons using different storage times~\cite{Picker05,Ezhov05}. The collaboration anticipates that a measurement with 0.1\,s uncertainty is possible. There are also ideas to use a quadrupole trap~\cite{Bowman05a} and to extract UCN into an Ioffe-type magnetic trap~\cite{Zimmer05b}.

A third approach to measuring \taun\ exists that, in principle, avoids this set of systematic problems. The most natural way to measure exponential decay is to acquire an ensemble of radioactive species and register the decay products. One can then simply fit the resulting time spectrum for the slope of the exponential function. With advances in the density of stored ultracold neutrons and magnetic field technology, a number of groups propose to eliminate the losses from wall scattering in material bottles and trap the neutrons in a combination of magnetic field gradients and gravity~\cite{Abov00}. In this way the neutrons never encounter a dissipative interaction during the measurement and the loss of neutrons from the trap should be caused only by neutron decay.   This class of experiments that continuously monitor neutron decays must become statistically competitive to make a significant statement, but the prospects are encouraging.

One experiment that has made significant progress uses ultracold neutrons that are magnetically confined in superfluid $^{4}$He~\cite{Huffman01}.   The UCNs fill a magnetic trap through the inelastic scattering of 0.89\,nm neutrons in superfluid \Hef\ (the superthermal process).  The decay electrons are registered via scintillations in the helium thus allowing one to directly fit for the exponential decay of the trapped neutrons.   The first result is in agreement with the currently accepted value of the free neutron lifetime, but the statistical uncertainty is large (60\,s)~\cite{Dzhosyuk05}. Upgrades to the apparatus are in progress to increase the number of trapped neutrons. The collaboration anticipates that a statistical precision of a few seconds will be possible in the near future.

\subsubsection{Radiative Neutron Decay}\label{sec:radiativedecay}

Beta decay of the neutron into a proton, electron, and antineutrino is accompanied by the emission of an inner-bremsstrahlung (IB) photon. While IB has been measured in nuclear beta decay and electron capture decays, it has never been observed in free neutron beta decay until very recently. Gaponov and Khafizov calculated the photon energy spectrum and branching ratio within a quantum electrodynamics (QED) framework~\cite{Gaponov96a,Gaponov96b,Gaponov00}, while Bernard {\it et al.} have calculated those properties as well as the photon polarization using heavy baryon chiral perturbation theory (HB$\chi$PT) including explicit $\Delta$ degrees of freedom~\cite{Bernard04a,Bernard04b}.  The QED calculation takes into account electron IB while the HB$\chi$PT calculation includes all terms to order $1/M$ (where $M$ is the nucleon mass) including photon emission from the effective weak vertex. These additional terms contribute at the level of less than 0.5\,\% and create only a slight change in the photon spectrum and branching ratio.  Both the photon energy spectrum and the photon polarization observables are dominated by electron IB. 

The experimental challenge was to distinguish the very low rate of radiative decay events in the large photon background of a neutron beam. The branching ratio above 15 keV is only about $3\times 10^{-3}$, which (coupled with the long neutron lifetime) makes the rate of detectable photons quite small. The first experiment to search for this decay mode resulted in an upper limit of $6.9\times10^{-3}$ (at the 90\,\% confidence level) for the branching ratio of photons between 35 keV and 100 keV~\cite{Beck02}. A more recent effort was able to definitively detect the radiative photons and measure the branching ratio at the 10\,\% level~\cite{Nico06}.  In the experiment, photons with energies between 15\,keV and 340\,keV were detected by a scintillating crystal coupled to an avalanche photodiode and were distinguished from uncorrelated background photons by coincidence with both the decay electron and proton~\cite{Fisher05}. Correlated background from external bremsstrahlung generated in the electron detector was estimated to contribute less than 3\,\% of the radiative decay event rate. The branching ratio was measured to be $(3.13\pm0.34)\times 10^{-3}$ and is consistent with the theoretical predictions of $2.85\times10^{-3}$ in the same energy region. Only a small fraction of the solid angle for photon detection was utilized in the experiment. A new detector is under construction that should permit a precision measurement of the photon energy spectrum and the branching ratio at the few percent level. A measurement below the 0.5\,\% level could reveal direct emission from the weak vertex. Furthermore, a measurement of the photon circular polarization could reveal information about the Dirac structure of the weak current~\cite{Gaponov00,Bernard04a}. Access to the radiative photon opens the possibility of new areas of investigation for neutron beta decay.

\subsection{Angular Correlation Experiments}\label{sec:Correlations}

\subsubsection{Spin-electron Asymmetry $A$}\label{sec:BigA}

With the neutron lifetime and one of the correlation coefficients $a$, $A$, or $B$, one can determine values for \ga\ and \gv. Because it has the greatest sensitivity to $\lambda$ and is more accessible experimentally, the spin-electron asymmetry $A$ has been measured more frequently and with greater precision.  Four independent measurements used in the PDG evaluation are not in good agreement with each other, so the PDG uses a weighted average for the central value and increases the overall uncertainty by a scale factor of 2.3~\cite{Yao06}. We discuss the two more recent measurements, one using a time projection chamber and one using an electron spectrometer, and the prospects for future improvement.

In the experiment of Schreckenbach et al.~\cite{Liaud97}, a beam of polarized cold neutrons was surrounded by a time projection chamber (TPC). Decay electrons passed through the drift chamber and were incident on plastic scintillators. The drift chamber recorded the ionization tracks in three dimensions while the scintillator gave the electron energy and start signal for the drift chamber. The TPC  provided good event identification and  reduced gamma ray backgrounds.  The result was $A=-0.1160\pm 0.0015$~\cite{Schreckenbach95}. The contributions to the overall uncertainty were roughly split between statistical and systematic uncertainties with the largest systematic contribution coming from the background subtraction.

The PERKEO II experiment also used a beam of cold polarized neutrons, but the decay electrons were extracted using a superconducting magnet in a split pair configuration. The field was transverse to the beam, so neutrons passed through the spectrometer but electrons were guided by the field to one of two scintillator detectors on each end. This arrangement had the advantage of achieving a $4\pi$ acceptance of electrons. They form an asymmetry from the electron spectra in the two detectors as a function of the electron energy; the difference in those quantities for the two detectors is directly related to the electron asymmetry. Their run produced $A=-0.1178\pm 0.0007$, where the main contributions to the uncertainty were in the neutron polarimetry, background subtraction, and electron detector response~\cite{Abele02}. 

The next version of PERKEO will use the new ballistic supermirror guide at the ILL with four times the fluence rate~\cite{Hase02}. They  intend to use a new configuration of crossed supermirror polarizers to make the neutron polarization more uniform in phase space~\cite{Petoukhov03,Kreuz05a}. The beam polarization can also be measured with a completely different method using an opaque \Het\ spin filter~\cite{Zimmer00b}. They anticipate reducing the main correction and uncertainty in the polarization analysis from 1.1\,\% to less than 0.25\,\% with an uncertainty of 0.1\,\% in that value. 

There are several other efforts underway to perform independent measurements of the electron asymmetry. The UCNA collaboration has made progress toward measuring $A$ using a superconducting solenoidal spectrometer~\cite{Young01}. UCN are produced in a solid deuterium moderator at LANSCE and transported to the spectrometer using diamond-coated guides. In the spectrometer, one produces highly polarized ($>99.9$\.\%) neutrons by passing them through a 6\,T magnetic field and into an open ended cylinder which increases the dwell time of the polarized UCN in the decay spectrometer. The decay electrons will be transported along the field lines to detectors at each end of the spectrometer. The detectors will initially consist of multiwire proportional counters backed by plastic scintillator, but other detectors are under development. Extensive measurements of electron backscattering on electron detector materials have been performed~\cite{Martin03,Martin06} The collaboration believes that a 0.2\,\% measurement is possible with three weeks of running.

Two other groups propose measuring the $A$ coefficient at the $10^{-3}$ level. The detector designs allow the possibility of measuring other decay correlation coefficients with the same apparatus. A group at PNPI is working on a magnetic spectrometer to be used with a highly collimated cold neutron beam~\cite{Serebrov05b}. The field guides decay particles to an electron detector at one end and a proton detector at the other. Their neutron polarimeter agrees with \Het -based spin filter methods at the $\approx 2\times 10^{-3}$ level~\cite{Serebrov95}.

The abBA collaboration proposes to use an electromagnetic spectrometer that guides both decay electrons and protons to detectors at each end of the spectrometer~\cite{Wilburn01}. The detector would be able to measure $a$, $A$, $B$, and the Fierz interference term $b$  all in the same apparatus. The detectors would be large-area segmented silicon detectors with thin entrance windows that allow the detection of both the proton and electron. The ability to detect coincidences greatly suppresses backgrounds and allows the measurement of residual backgrounds. The magnetic field guides the decay products to conjugate points on the segmented Si detectors and provides $4\pi$ detection of both electrons and protons and suppression of backgrounds by use of coincidences.  The apparatus is being designed for use at a pulsed spallation source to exploit background reduction and perform neutron polarimetry. The neutrons can be polarized by transmission through polarized \Het, whose spin-dependent absorption cross section possesses an accurately-known neutron energy dependence that can be exploited for accurate neutron polarization measurement~\cite{Coulter88,Coulter90,Greene95, Zimmer00b,Rich02,Penttila05, Wietfeldt05a}.

One strategy to eliminate the need for absolute neutron polarization measurements is to measure both $PA$ and $PB$ simultaneously in the same apparatus ($P$ is the neutron polarization) and take the ratio. From this ratio $\lambda$ can be calculated directly. Many of the planned experiments propose to implement this method, which was first performed in the experiment of Mostovoi~\cite{Mostovoi01} with the result $\lambda=-1.2686 \pm 0.0046$ 

\subsubsection{Spin-antineutrino Asymmetry $B$}\label{sec:BigB}

The electron asymmetry and antineutrino asymmetry provide complementary information. \ga\ is equal to $-1$ in the SM Lagrangian at the quark level but is renormalized in hadrons by the strong interaction.  Since  \ga\ is nearly $-1$, $A$ is close to zero, and $B$ is near unity, $B$ is not particularly sensitive to $\lambda$ but is more sensitive to certain non-SM extensions, such as extended left-right symmetric models.

Left-right symmetric models, which are motivated in part to restore parity conservation at high energy scales, add a new right-handed charged gauge-boson $W_2$ with mass $M_2$ and four new parameters to be constrained by experiments: a mixing angle $\zeta$, $\delta=(M_1/M_2)^2$, $r_g = g_R/g_L$ the ratio of the right- and left-handed gauge coupling strengths $g_R$ and $g_L$, and $R_K=V^R_{\text{ud}}/V^L_{\text{ud}}$ where $R$ and $L$ designate  the right and left sectors. In the manifest left-right symmetric model (MLRM), $r_g=r_K=1$, and in the SM $\delta=0$.

The best direct search for $W_2$ sets a lower limit of 652\,GeV/c$^2$ on $M_2$~\cite{Abe95}, and the mass limit on a right-handed vector boson comes from muon decay and is 406\,GeV/c$^2$~\cite{Jodido86}. In the MLRM where there are only two parameters ($\zeta$ and $\delta$), constraints from other systems are better than the neutron constraints. For the extended left-right model, however, neutron-derived constraints are complementary to the other searches. Another area in which to search for right-handed currents is the decay of the neutron into a hydrogen atom and antineutrino since one of the hyperfine levels of hydrogen in this decay mode cannot be populated unless right-handed currents are present~\cite{Byrne01}. The small branching ratio has precluded a search so far.

In the last three decades, there have been only two new measurements of the antineutrino asymmetry. Since the antineutrino cannot be conveniently detected, its momentum was deduced from electron-proton coincidence measurements.  Electrons from the decay of polarized neutrons were detected by plastic scintillators, and protons were detected by an assembly of two microchannel plates. From the electron energy and proton time-of-flight, one can reconstruct the antineutrino momentum. The first  measurement was carried out at PNPI and produced a result of $B=0.9894\pm 0.0083$~\cite{Kuznetzov95}. A second run at the ILL used largely the same apparatus and measured $B=0.9821\pm 0.0040$~\cite{Serebrov98}, where the largest reduction in the overall uncertainty came from improved statistics.

A recent measurement of $B$ was performed using the PERKEO II apparatus~\cite{Kreuz05b}. Typically, one detects electron-proton coincidences using one detector for each particle. The PERKEO II measurement uses two detectors, one in each hemisphere of the detector, that can detect both electrons and protons. Electrons are detected using plastic scintillator, while the protons are accelerated on a thin carbon foil placed in front of the scintillator. The resulting secondary electrons are guided onto the
electron detectors. This technique reduces systematics and increases the sensitivity to $B$. The result, $B=0.967 \pm 0.006 (stat) \pm 0.010 (sys)$ can be further improved in PERKEO III.  The PERKEO II apparatus also lends itself to the measurement of the proton asymmetry coefficient $C$, and the first measurement of $C$ has recently been reported~\cite{Abele05b}.

\subsubsection{Electron-antineutrino Asymmetry $a$}\label{sec:Littlea}

Although the electron-antineutrino asymmetry $a$ has approximately the same sensitivity to $\lambda$ as $A$, it is only known to 4\,\%. Since 1978~\cite{Stratowa78}, there has been only one new measurement. Unlike $A$, its measurement does not require polarized neutrons. The experimental difficulty lies with the energy measurement of the recoil protons, whose spectral shape is slightly distorted for nonzero $a$. A precision measurement of $a$ would produce an independent measurement of $\lambda$, an improved test of CKM unitarity, and model-independent tests of new physics. The values of $a$, $A$, and $B$ can be related to the strength of hypothetical right-handed weak forces and scalar and tensor forces~\cite{Dubbers91b,Yerozolimsky91}, and it was recently shown that a precise comparison of $a$ and $A$ can place stringent limits on possible conserved-vector-current (CVC) violation and second class currents in neutron decay~\cite{Gardner01}.

The most recent determination of $a$ comes from measurements of the integrated energy spectrum of recoil protons stored in an ion trap~\cite{Byrne02, Byrne05b}. A collimated beam of cold neutrons passed though a proton trap consisting of annular  electrodes coaxial with a magnetic field whose strength varied from 0.6\,T to 4.3\,T over the length of the trap. Protons created inside the volume were trapped, and those created in a high field region were adiabatically focused onto a mirror in the low field region. The trap was periodically emptied and the protons counted as a function of the mirror potential. The result of $a=-0.1054\pm 0.0055$ is in good agreement with the previous measurement and of comparable precision.

The precision of  $a$ measurements must be improved to the level of $A$ experiments to constrain  $\lambda$. There are two major efforts underway to improve the precision of $a$,  $a$SPECT~\cite{Zimmer00c} and aCORN~\cite{Wietfeldt05b, Collett05}. aCORN relies on the measurement of an asymmetry in the coincidence detection of electrons and recoil proton that is proportional to $a$~\cite{Yerozolimsky04}. The asymmetry is formed by carefully restricting the phase space for the decay in a magnetic spectrometer so that decay events with parallel and antiparallel electron and antineutrino momenta are separated in the coincidence timing spectrum. $a$ is directly proportional to the relative number of events, and there is no need for precise spectroscopy of the low energy protons. The experiment will be built and tested at the Low Energy Neutron Source (LENS)~\cite{Baxter03, Leuschner05} and then run at NIST where a measurement of approximately 1\,\% accuracy is feasible.

In the $a$SPECT experiment, one again measures a proton energy spectrum as a function of a potential, similar to the idea used for the proton trap experiment.  One increases the statistical power by completely separating the source part and the spectroscopy part of the apparatus. A cold neutron beam will pass through a region of strong, homogeneous magnetic field transverse to the beam. The decay protons with initial momentum component along the field direction will be directed toward a detector. Near the detector is a region of weaker magnetic field and electrostatic retardation potentials, and only those protons with sufficient energy to overcome the barrier continue on to the detector. Registering the protons as a function of the retardation potential gives the recoil proton spectrum, which one fits to extract $a$. This type of integral spectrometer is similar in spirit to the latest electron spectrometers used in tritium beta decay searches for the electron neutrino mass. The collaboration believes that a statistical uncertainty of approximately 0.25\,\% is achievable~\cite{Gluck05} and has commenced testing the apparatus at the ILL. Other ideas for measuring $a$ have recently been proposed~\cite{Bowman05b}.

\section{SEARCHES FOR NONSTANDARD $T$ AND $B$ VIOLATION}\label{sec:TViolation}

The physical origins of the observed $CP$ violation in nature, first seen in the neutral kaon system~\cite{Christenson64}, remain obscure.  $CP$ violation implies time-reversal symmetry $T$ violation (and vice versa) through the $CPT$ theorem. Recent experiments have reported measuring $CP$ violation in the $K^o \to 2\pi$ amplitudes~\cite{Alavi99,Fanti99} and in the decays of the neutral $B$-mesons~\cite{Aubert02,Abe02}.  The Standard Model can accommodate the possibility of $CP$ violation through a complex phase $\delta_{\text{KM}}$ in the CKM quark mixing matrix.  To date there is no firm evidence against the possibility that the observed $CP$-violation effects are due to this phase~\cite{Herczeg05}, but the question remains whether or not there are sources of $CP$-violation other than $\delta_{\text{KM}}$. There is indirect evidence for this possibility from cosmology; it appears that  $\delta_{\text{KM}}$ is not sufficient to generate the baryon asymmetry of the universe in the Big Bang model.  One area to probe for the existence of new $CP$-violating interactions is systems involving first-generation quarks and leptons for which the contribution from $\delta_{\text{KM}}$ is typically suppressed.  Examples of observables of this kind are electric dipole moments of the neutron, leptons, and atoms~\cite{Khriplovich96a} and $T$-odd correlations in leptonic and semileptonic decays and reactions and neutron spin-dependent transmission asymmetries. 

\subsection{EDM Theoretical Framework}\label{sec:TimeThy}

\begin{figure}
\begin{center}
%\epsfscale1200         % Figure enlarged to 120 (MAC)%
%\epsfxsize10pc         %
%\centerline{\epsfbox{fig3.eps}}
\includegraphics[width=4.5in]{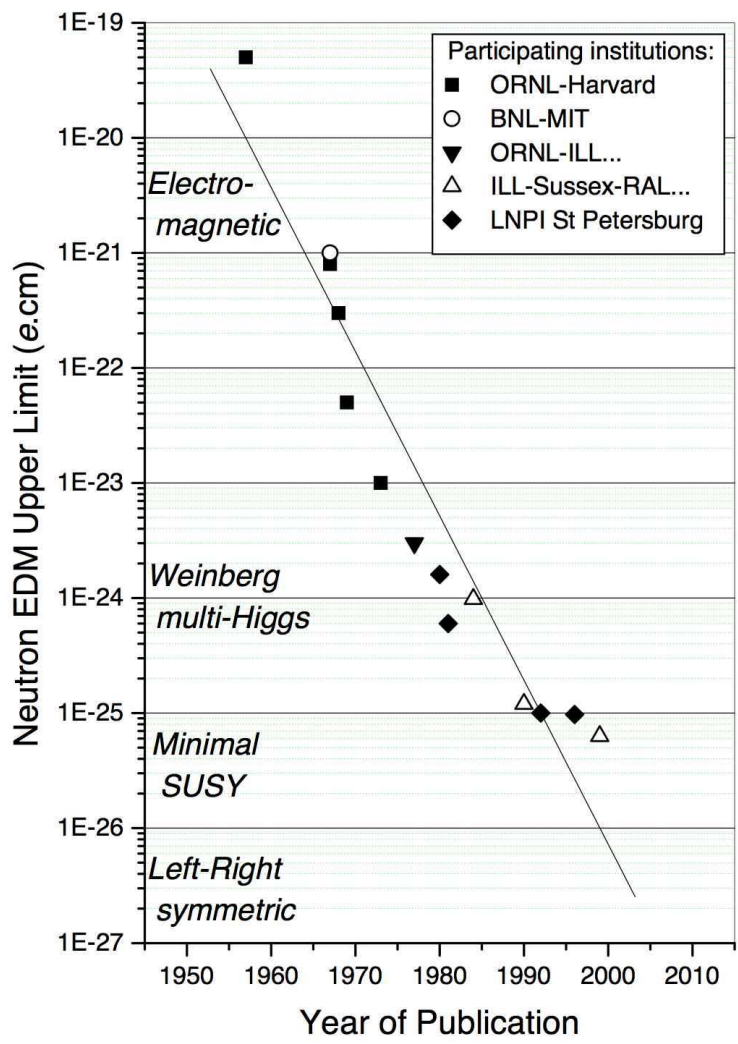}
\caption{Plots showing the history of neutron EDM limit and some of the ranges for different theoretical predictions. (Plot courtesy of P. Harris, University of Sussex.)}
\label{fig:edm}
\end{center}
\end{figure}

The search for the neutron electric dipole moment addresses issues which lie at the heart of modern cosmology and particle physics. The current limit on the permanent EDM of the neutron represents one of the most sensitive null measurements in all of physics and has eliminated many theories and extensions to the Standard Model (see Figure~\ref{fig:edm}).  The reader is directed to  comprehensive reviews of EDM experiments in Refs.~\cite{Ramsey82, Ramsey90, Pendlebury93, Golub94}.

The energy of a neutral spin-1/2 particle with an electric dipole moment $d_n$ in an electric field $\vec{E}$ is  $E_n = d_n\vec{\sigma}\cdot\vec{E}$, where   $\vec{\sigma}$ is the Pauli spin matrix.  This expression is odd under $T$ and $P$. The current experimental bound on the neutron EDM is $d_n< 0.63 \times 10^{-25}$\,\ecm\, (90\,\% CL)~\cite{Harris99,Yao06}. In the SM, there are two sources of $CP$ violation. One source is the complex phase ${\delta_{KM}}$  in the CKM matrix.  The other source is a possible term in the QCD Lagrangian itself, the so-called $\theta$-term
\begin{equation}
\label{eq:QCD}
\mathcal{L}_{QCD} = \mathcal{L}_{QCD,\theta=0}+{{\theta {g_s}^2}\over 32 \pi^2}G_{\mu\nu}\tilde{G}^{\mu\nu},
\end{equation}
which explicitly violates $CP$ symmetry because of the appearance of the product of the gluonic field operator $G$ and its dual $\tilde G$.  Since $G$ couples to quarks but does not induce flavor change, $d_n$ is much more sensitive to $\theta$ than it is to $\delta_{KM}$.  Thus, measurement of $d_n$  determines an important parameter of the SM.  Calculations have shown that $d_n\sim O(10^{-16}\theta)$\,\ecm~\cite{Baluni79,Crewther79a,Crewther79b}.

The observed limit on $d_n$ allows one to conclude that $\theta<10^{-(9\pm 1)}$~\cite{Bigi00}.  Since the natural scale is $\theta\sim O(1)$, the very small value for $\theta$ (known as the strong $CP$ problem) requires an explanation. One attempt augments the SM by a global U(1) symmetry (referred to as the Peccei-Quinn symmetry), whose spontaneous breakdown gives rise to Goldstone bosons called axions~\cite{Peccei77a,Peccei77b}.  The $\theta$-term is then essentially eliminated by the vacuum expectation value of the axion. No axions have yet been observed despite extensive searches. 

Since $CP$ violation through the phase in the CKM matrix involves flavor  mixing of higher generation quarks, $d_n$ is very small in the SM; calculations predict it to be $10^{-32}$\,\ecm\ to $10^{-31}$\,\ecm~\cite{Khriplovich82,Gavela82}, several orders of magnitude beyond the reach of any experiment being considered at present.  Models of new physics, including left-right symmetric models, non-minimal models in the Higgs sector, and supersymmetric (SUSY) models, allow for $CP$ violating mechanisms not found in the SM, including terms that do not change flavor.  Searches for electric dipole moments in the neutron, leptons, and atoms, which are particularly insensitive to flavor-changing parameters, can therefore strongly constrain such models.

\subsubsection{Baryon Asymmetry}\label{sec:BAsymmetry}

Antimatter appears to be rare in the universe, and there is an asymmetry between the number of baryons and antibaryons. Although the SM possesses a nonperturbative mechanism to violate $B$, no experiments have seen $B$ violation, and it is natural to speculate on the origin of the baryon asymmetry of the universe. There are two outstanding facts:  baryons make up only 5\% of the total energy density of the universe and the ratio of baryons to photons is very small. The ratio $n_{B}/ n_{\gamma}=(6.1 \pm 0.3) \times 10^{-10}$ is known independently both from BBN and fluctuations in the microwave background~\cite{Bennett03}. Sakharov first raised the possibility of calculating the baryon asymmetry from basic principles~\cite{Sakharov67}. He identified three criteria that, if satisfied simultaneously, will lead to a baryon asymmetry from an initial $B=0$ state:  baryon number violation, $CP$ violation, and departure from thermal equilibrium. One way to explain the asymmetry assumes that all three of these conditions were met at some very early time in the universe and that this physics will remain inaccessible to us, with the $B$ asymmetry effectively an initial condition. However, the scenario of inflation in the early universe - which generates a flat universe, solves various cosmological problems, and generates a spectrum of primordial density fluctuations consistent with observation - would dilute any such early $B$ asymmetry to a negligible level. In this case the $B$ asymmetry must be regenerated through subsequent post-inflationary  processes, and there is hope that it is calculable from first principles~\cite{Riotto99}.   

Although the SM contains processes that satisfy the first two conditions and the Big Bang certainly satisfies the third, it fails by many orders of magnitude in its estimate of the size of the baryon asymmetry. Grand unified theory (GUT) baryogenesis at $T \sim10^{29}$\,K corresponding to a mass scale on the order of $10^{16}$\,GeV is disfavored by inflation. Electroweak baryogenesis~\cite{Kuzmin85, Shaposhnikov87}, which relies on a nonperturbative $B-L$-violating mechanism present in the SM due to nonperturbative electroweak fields~\cite{tHooft76} combined with $CP$ violation and a departure from equilibrium at the electroweak phase transition, remains a viable possibility. Leptogenesis~\cite{Fukugita86, Buchmiller05}  combined with $B-L$ conserving processes to get the $B$ asymmetry and the Affleck-Dine mechanism are favored speculations at the moment~\cite{Dine04}. 

It  appears that some physics beyond the SM, including new sources of $CP$ violation that may lead to a measurable value for $d_n$, must exist if the observed baryon asymmetry is to be understood. The minimal supersymmetric extension of the SM (MSSM)~\cite{Riotto98} can possess small values of the $CP$-violating phases (consistent with constraints from $d_n$) that generate the baryon asymmetry. Within the broad framework of non-minimal SUSY models, including GUTs, there are numerous new sources of $CP$ violation in complex Yukawa couplings and other Higgs parameters that may have observable effects on the neutron EDM~\cite{Barbieri96,Dimopoulos95,Khriplovich96b}. The present limit on the neutron EDM already severely constrains many SUSY models. The question of the relationship between electric dipole moments and the baryon asymmetry generated at the electroweak scale is just beginning to be addressed in a quantitative manner~\cite{Musolf05} 

\subsection{Electric Dipole Moment Experiments}\label{sec:EDM}

EDM experiments monitor the possible electric field dependence of the precession frequency of polarized neutrons using  the well-known Ramsey interferometric technique of separated oscillatory fields. Static electric $E_0$ and magnetic $B_0$ fields are applied to the polarized neutrons. The neutron spin state is then governed by the Hamiltonian $H= -{\vec\mu}\cdot{\vec {B_0}}\pm{\vec {d_n}}\cdot{\vec {E_0}}$. An RF magnetic field of frequency $\omega_{a}$ is applied to tilt the neutron polarization normal to $E_{0}$ and $B_{0}$ and it starts to precess with a frequency $\omega_{R}$. After a free precession time $T$ a second tilt pulse in phase with the first is applied and the neutron polarization direction is proportional to $(\omega_{R}-\omega_{a})T$. The Larmor precession frequency of the neutron depends on the direction of the applied field $\vec{E_0}$ relative to $\vec\mu$. An EDM would appear as a change in $\omega_{R}$  as the electric field is reversed.

The two most stringent limits on $d_n$ come from Altarev et al. at PNPI~\cite{Altarev96} and Harris et al. at ILL~\cite{Harris99}. Both experiments used stored UCN. The PNPI apparatus contained two UCN storage chambers with oppositely-directed electric fields.   A nonzero EDM would cause frequency shifts of  opposite sign in each of the chambers, and some sources of magnetic field noise are suppressed with simultaneous measurements with both fields.  Nevertheless, slowing varying magnetic fields remained a significant source of systematic uncertainty. In the experiment of Harris et al., a polarized $^{199}$Hg  comagnetometer occupying approximately the same volume as the neutrons was introduced into the storage volume to continuously monitor the magnetic field.   The comagnetometer was essential in eliminating stray magnetic fields as a major source of systematic uncertainty. The experimental accuracy was limited by neutron counting statistics. 

There are ambitious efforts underway to improve the current neutron EDM limit by one to two orders of magnitude. All of the experiments attempt to increase the number of UCN, the observation time, and the size of the applied electric field. As the measurements become more sensitive more subtle systematic effects, such as a recently-discussed frequency shift associated with certain UCN trajectories in a magnetic field gradient~\cite{Pendlebury04, Golub05a}, must be considered. The CryoEDM collaboration intends to produce UCN through the superthermal process and transport them to a separate measurement chamber containing superfluid helium.  Liquid helium should allow electric field values that are several times larger than used in past experiments, and the cryogenically pure environment should permit longer UCN storage times. The collaboration proposes to use a multichamber spectrometer for compensation of field fluctuations by means of SQUID magnetometers.

The nEDM collaboration proposes to search for the neutron EDM using a double chamber  storage cell to suppress magnetic field fluctuations and allow one to extract $d_n$ from the simultaneous measurement of chambers with opposite electric field values. The magnetic field would be inferred from a set of laser optically pumped Cs magnetometers placed outside the storage cells. The UCN would come from the solid deuterium UCN source under construction at PSI.

Another EDM experiment under development~\cite{Golub94, Golub05b} also proposes to increase the UCN density using downscattering of UCN in superfluid \Hef\ and exploit the large electric fields achievable in helium. The experiment will use polarized \Het\ atoms as the comagnetometer in a bath of superfluid helium at a temperature of approximately 300\,mK. The comagnetometer should sample the same fields as the neutrons, and historically this has been a very important check on potential systematic errors from magnetic field fluctuations. The strong spin dependence of the \Het\ neutron absorption cross section allows the relative orientation of the neutron and \Het\ spins to be continuously monitored through the intensity of the scintillation light in the helium in addition to SQUID magnetometry techniques. The polarized \Het\ slowly depolarizes in the storage chamber and must be periodically replenished: the low temperature dynamics of \Het\ motion in superfluid \Hef\ makes this possible, and unpolarized test measurements have been performed~\cite{Lamoreaux02, Hayden04}.  One obtains $d_n$ by measuring the difference in the neutron and \Het\ precession frequencies for the different orientation of the electric  field, which can be measured optically in superfluid helium using the Kerr effect~\cite{Sushkov04}. 

There are also preparations underway to search for the neutron EDM using dynamical diffraction from noncentrosymmetric perfect crystals. In dynamical diffraction the incident neutron plane wave state $|k>$ is split as it enters the crystal into two coherent branches $|k_{+}>$ and $|k_{-}>$ with slightly different momenta and energies. The probability density of these two states in the crystal is concentrated along and in between the lattice planes, respectively. In noncentrosymmetric crystals the position of the electric field maxima can be displaced with respect to the nuclei. Therefore one of the branches can experience  interplanar electric fields($\sim 10^{9}$V/cm) which are orders of magnitude larger than can be achieved through application of external fields in vacuum or in isotropic media~\cite{Voronin00,Zeyen00}. The presence of a neutron EDM would produce an extra relative phase shift between the two interfering branches. Such experiments must contend with potentially large systematic effects such as those from neutron spin-orbit scattering from the atoms~\cite{Golub99}. A number of experiments which investigate neutron optical issues relevant for an eventual EDM measurement of this type have been performed recently~\cite{Voronin01,Dombeck01,Fedorov02a, Fedorov02b,Fedorov03, Voronin03, Fedorov05}. 

\subsection{$T$-violation in Neutron Beta Decay}

With its small SM values of time-reversal violating observables, neutron beta decay also provides a excellent laboratory in which to search for $T$ violation. Leptoquark, left-right symmetric models, and exotic fermion models can all lead to violations of time-reversal symmetry at potentially measurable levels~\cite{Herczeg01}. One possible $T$-odd correlation in polarized  neutron decay is 
$D \vec{\sigma}_{n} \cdot \left({\vec{p}_{e} \times \vec{p}_{p}} \right)$, where $\vec{p}_{p}$ is the momentum of the recoil proton.   The $D$ coefficient is sensitive only to $T$-odd interactions with vector  and axial vector currents. In a theory with such currents, the coefficients of the correlations depend on the magnitude and phase of
$\lambda=|\lambda|e^{-i\phi}$.

$D$ has $T$-even contributions from phase shifts due to pure Coulomb and weak magnetism scattering. The Coulomb term vanishes in lowest order in V-A theory~\cite{Jackson57}, but scalar and tensor interactions could contribute. Fierz interference coefficient measurements~\cite{Wenninger68, Carnoy94} can be used to limit this possible contribution to $|D^{EM}| < (2.8\times 10^{-5}){m_e\over p_e}$. Interference between Coulomb scattering amplitudes and the weak magnetism amplitudes produces a final state effect of order ${E_e}^2/{p_em_n}$. This weak magnetism effect is predicted to be $|D^{WM}| =1.1\times 10^{-5}$~\cite{Callan67}. References~\cite{Herczeg01, Herczeg05} summarizes the current constraints on $D$ from analyses of data on other $T$-odd observables for the SM and extensions.

The EDM violates both $T$ and $P$ symmetries, whereas a $D$ coefficient violates $T$ but conserves $P$. This makes the two classes of experiments sensitive to different SM extensions.  Although constraints on $T$-violating, $P$-conserving interactions  can be derived from EDM measurements, these constraints may be model dependent~\cite{RamseyM01}, and  EDM and neutron decay searches for $T$ violation are complementary.

\subsection{$D$- and $R$-coefficient Measurements}\label{sec:DCoeff}

In the last decade, there have been two major experimental efforts, \EMIT\ and Trine, to improve the limit on the $D$ coefficient in neutron decay.  Each requires an intense, longitudinally polarized beam of cold neutrons around which one places  proton and electron detectors alternating in azimuth. Coincidence data are collected in electron-proton pairs as a function of the neutron spin state to search for the triple correlation.

In the \EMIT\ experiment, the detector consisted of four electron detectors and four proton detectors arranged octagonally around the neutron beam~\cite{Lising00}. The octagonal geometry maximized the experiment's sensitivity to $D$ by balancing the sine dependence of the cross product $\vec{\sigma}_{n} \cdot \left({\vec{p}_{e} \times \vec{p}_{p}} \right)$ with the large angles between the proton and electron momenta that are favored by kinematics. The decay protons drifted in a field free region before being focused by a 30\,kV to 37\,kV potential into an array of PIN diode detectors.  With its maximum recoil energy of 750\,eV, most of the protons arrived approximately 1\,$\mu$s after the electrons.   Detector pairs were grouped in the analysis to reduce potential systematic effects from neutron transverse polarization.  The result from the first run of \EMIT\  yielded an improved limit of $D=[-0.6\pm 1.2{\rm (stat)} \pm 0.5{\rm (sys)}]\times 10^{-3}$~\cite{Lising00}.

Currently, the best constraint on $D$ comes from the Trine collaboration, which reports $D=[-2.8\pm 6.4{\rm (stat)} \pm 3.0{\rm (sys)}]\times 10^{-4}$~\cite{Soldner04}. They used two proton detectors and two electron detectors in a rectangular geometry. The proton detectors were comprised of arrays of thin-window, low-noise PIN diodes. The detectors were held at ground while the neutron beam was set to a potential of 25\,kV by surrounding it with a high voltage electrode; the field was shaped to focus the decay protons onto the PIN arrays. The electrons were detected by plastic scintillators in coincidence with multi-wire proportional chambers. This coincidence provides reduction in the gamma-ray background rates and positional information on the decay, thus minimizing some sources of systematic uncertainty. 

The current PDG evaluation obtains a new value for the neutron $D$ coefficient of  $(-4\pm 6)\times 10^{-4}$, which constrains the phase of \ga/\gv\ to $180.06^\circ\pm 0.07^\circ$~\cite{Yao06}.
Neither experiment produced a statistically limited result, and  both collaborations upgraded their detectors and performed second runs~\cite{Mumm04, Plonka04}. In the near future it is reasonable to anticipate new results that will put a limit on $D$ very near $10^{-4}$. Although there have been discussions and ideas for experiments using UCNs, there are currently no concrete proposals to further improve the limit on $D$.

Another $T$-odd correlation that may be present in neutron decay is the $R$ correlation, $R \vec{\sigma_n} \cdot \left({\vec{\sigma_{e}} \times \vec{p}_{e}} \right)$, where $\sigma_n$ is the neutron spin and $\vec{\sigma_{e}}$ is the spin of the decay electron. A nonzero $R$ requires the presence of scalar or tensor couplings and is sensitive to different SM extensions than $D$. An effort is underway at PSI to measure $R$ in the neutron by measuring the neutron polarization and the momentum and transverse polarization of the decay electron at the level of $5\times 10^{-3}$~\cite{Bodek05}. 

Neutron decay is a mixed Fermi and Gamov-Teller decay, so a measurement of $R$  would produce a limit on both scalar and tensor $T$-odd couplings. The limit on $R$ achieved in  $^8$Li Gamov-teller decay  of $R=(0.9\pm 2.2)\times10^{-3}$ now sets the most stringent limits for time-reversal violating tensor couplings in semileptonic weak decays, $-0.022< Im(C_T+C^{\prime}_T)/C_A<0.017$~\cite{Huber03}.
 
 \subsection{$T$ Violation in Neutron Reactions}\label{sec:Toptics}

$T$ violation can lead to terms in the forward scattering amplitude for polarized neutrons in polarized or aligned targets of the form $\vec{s} \cdot \left({\vec{k} \times \vec{I}} \right)$ and  ${\vec s} \cdot ({\vec k} \times {\vec I})({\vec k} \cdot {\vec I})$~\cite{Stodolsky82}, where $\vec{s}$ is the neutron spin and $\vec{I}$ is the nuclear polarization. Since the  enhancement mechanisms for parity violation in compound resonances of heavy nuclei are also applicable to $T$-odd interactions~\cite{Bunakov82, Flambaum95, Bunakov97} (see Section~\ref{sec:NNInteractions}), it is possible for such searches to reach interesting levels of sensitivity. Although in principle $T$-odd observables in forward scattering are motion-reversal invariant and therefore not subject to final state effects, in practice the large spin dependence of the neutron-nucleus strong interaction in a polarized target can induce large potential sources of systematic errors which require careful study~\cite{Lamoreaux94}.

These systematic effects are smaller in aligned targets, and a search for the $T$-odd fivefold correlation in MeV polarized neutron transmission in an aligned holmium target has set the best direct limit on such interactions~\cite{Huffman97}. This $P$-even, $T$-odd correlation is especially interesting to search for, since there exists no renormalizable gauge theories with $P$-even $T$-odd tree-level gauge boson couplings between quarks~\cite{Herczeg92}. Although EDM limits can also be used to constrain $P$-even $T$-odd interactions in many models, in general only direct measurements can set model-independent bounds~\cite{RamseyM99}. A JINR-ITEP collaboration is  preparing to perform a search for the $P$-even $T$-odd fivefold correlation with low energy neutrons on p-wave resonances using microwave-induced dynamical nuclear alignment to order the nuclei~\cite{Atsarkin00, Barabanov03, Barabanov05}. A shift in the zero crossing of the forward-backward asymmetry in unpolarized neutron capture on p-wave resonances has also been considered as a signal for $P$-even, $T$-odd interactions~\cite{Barabanov93} but it is not a null test and although the statistical precision of the bound from $^{113}$Cd is at the $10^{-4}$ level, one must accumulate a number of results either in different nuclei or in different resonances in the same nucleus in order to construct a statistically meaningful bound from such measurements~\cite{Davis99}.
 
Searches for the 3-fold $P$-odd $T$-odd correlation require a polarized target. Nuclei have been identified ($^{139}$La,$^{131}$Xe) that are polarizable in macroscopic quantities and possess large parity-odd asymmetries at low energy p-wave resonances~\cite{Alfimenkov82a, Szymanski96, Skoy96}. The first steps toward such an experiment are in progress at KEK~\cite{Masuda00, Skoy05, Loukachevitch05, Masuda05}. Polarized neutron capture on unpolarized targets can also set limits on $P$-odd, $T$-odd interactions~\cite{Flambaum84}, but because there are inelastic reactions, they are susceptible to initial and final state effects that can mimic $T$ violation~\cite{Davis04, Davis05}.  
 
\subsection{Neutron-antineutron Oscillations}\label{sec:NNbar}

An observation of neutron-antineutron oscillations would constitute a discovery of fundamental importance~\cite{Kuzmin70}. It requires a change of baryon number by 2 units and no change in lepton number and therefore must be mediated by an interaction outside the SM of particle physics.
Among neutral mesons with sufficiently long lifetimes (kaons, $B$ mesons) ) or neutral systems with flavor changing processes (neutrinos), oscillations are no longer a surprising phenomenon. The observation of oscillations in these systems has yielded information on aspects of physics (lepton number violation, $T$ violation, neutrino mass) that are not accessible using less sensitive techniques.  It is reasonable  to hope that a search for oscillations in the neutron, the only neutral baryon which is sufficiently long-lived to conduct a practical experiment, may uncover new processes in nature.

In the SM there are no (perturbative) renormalizable interactions one can write down which violate $B$, and any nonrenormalizable operator that can induce $B$ violation must be naively suppressed by some heavy mass scale.  The effective operator for neutron-antineutron oscillations involves a dimension 9 operator to change the 3 quarks in the neutron into 3 antiquarks and is suppressed by some mass scale to the 5th power.  Some SM extensions lead to $B$ violation by 2 units and not 1 unit. Examples include left-right symmetric models~\cite{Mohapathra80}  with a local $B$-$L$ symmetry needed to generate small Majorana neutrino masses by the seesaw mechanism, SUSY models with spontaneously broken $B-L$ symmetry~\cite{Babu01}, and theories with compactified extra dimensions which attempt to solve hierarchy problems by introducing a much lower scale (TeV) for the onset of quantum gravity~\cite{Dvali99}. In these  cases proton decay is unobservably small but neutron-antineutron oscillations can occur close to the present limit. A general analysis of all operators with scalar bilinears that couple to two SM fermion fields uncovers operators which can only lead to neutron-antineutron oscillations and not to proton decay~\cite{Klapdor02}.

The last experiment in the free neutron system at the ILL set an upper limit of $8.6\times 10^7$\,s (90\,\% confidence level) on the oscillation time~\cite{BaldoCeolin94}. Neutrons from the ILL cold source were extracted using a diverging neutron guide that compressed the transverse phase space of the neutron beam. The neutrons traversed a magnetically-shielded vacuum chamber of about 100 meters before striking a carbon foil surrounded by an antineutron detector. The residual magnetic field was measured in-situ by polarizing the neutrons and looking for the rotation of the plane of polarization using spin-echo techniques.  No antineutron events were observed.   Translated into a mass scale, this limit excludes mass scales for the effective operator that induces oscillations below $\sim 100$\,TeV. 
A similar indirect limit is set by the absence of evidence for spontaneous neutron-antineutron oscillations in nuclei in large underground detectors built for proton decay and neutrino oscillation studies~\cite{Kamyshkov03}.  Although there has been some discussion of possible strategies to improve on the bounds from direct searches using cold and ultracold neutrons~\cite{Kamyshkov02,nnbar02}, there are no new free neutron-antineutron oscillation searches underway.

The neutron and antineutron are measured to have the same mass to a precision of only $10^{-4}$. If CPT symmetry is somehow violated and the neutron and antineutron mass are different, then neutron-antineutron oscillations are greatly suppressed~\cite{Lamoreaux91}.

There are many other modes one might imagine for $B$-violating processes involving neutrons. Of the experimental limits for the many possible processes, the poorest limits are those for so-called ``invisible'' modes of neutron decay, favored in some models~\cite{Mohapathra03} in which one or two neutrons decay to neutrinos. If such processes were to occur in nuclei, they would leave neutron holes whose deexcitation would leave characteristic signatures. Recent data from underground detectors built to observe neutrino oscillations place much stronger constraints on such n and nn decay modes~\cite{Ahmed04, Back03, Araki05}.    

\section{NEUTRON-NUCLEON WEAK INTERACTIONS}\label{sec:NNInteractions}

The most obvious consequence of the weak interaction for neutrons is that it makes neutrons unstable. In addition to the coupling of quarks to leptons that allows neutron to decay, electroweak theory also predicts (and experiments confirm) that there are weak interactions between the quarks in the neutron with couplings comparable in size to those involved in neutron decay. The weak nucleon-nucleon (NN) interaction is a unique probe of strongly interacting systems. This section presents an overview of the importance of NN interactions for QCD and the status of the experimental efforts. Reviews of various aspects of the field can be found in References~\cite{Krupchitskii94, Haeberli95,Desplanques98,Dmitriev04, Flambaum95,  Holstein04}.

\subsection{Overview}\label{sec:NNMotivation}

The dynamics of the quarks in the nucleon are dominated by momentum transfers that are less than that set by the QCD scale of 1\,GeV/c. In this regime QCD becomes so strong that quarks are permanently confined, and therefore the quark-quark weak interactions appear  through the NN weak interactions that they induce.  At these energies quark-quark weak amplitudes are of order $10^{-7}$ of strong amplitudes primarily because of the short range of the quark weak interactions through $W$ and $Z$ exchange. 

Assuming that it is correctly described by the electroweak theory at low energy scales, the quark-quark weak interaction can be used as an internal probe of strongly interacting systems. Collider measurements have verified the SM predictions for quark-quark weak couplings for large momentum transfers at the 10\,\% level~\cite{Arnison86}. The weak amplitudes are too weak to significantly affect the strong dynamics or to excite the system, and therefore it probes quark-quark correlations in the QCD ground state.  The effects of the quark-quark weak interaction can be isolated from the strong interaction using  parity violation. The short range of the quark-quark weak interaction and its ability to violate parity make it visible and sensitive to interesting aspects of strongly interacting systems, as seen in four cases.

1) The ground state of the strongly interacting limit of QCD is a problem of fundamental importance. Although the dynamics that lead to the spontaneous breakdown of chiral symmetry in QCD are not yet understood, one of the leading models assumes the importance of fluctuating nonperturbative gluon field configurations called instantons~\cite{Belavin75}. They induce four-quark vertices that flip the quark helicity and localize the quark wave function through a mechanism similar to Anderson localization of electrons in disordered metals~\cite{Schafer98, Diakonov03}.   Some aspects of QCD spectroscopy and the high density limit can be understood by assuming that quark-diquark configurations in the nucleon are important~\cite{Anselmino93}. The mechanism for the phenomenon of color superconductivity in the high density limit of QCD  consists of a BCS-like condensation of diquarks~\cite{Alford99}. The quark-quark weak interaction in the nucleon in the low energy limit induces four-quark operators with a known spin and flavor dependence whose relative sizes are in principle sensitive to these and other correlation phenomena in the ground state of QCD.  

2) With experimental information on the low energy parity-violating (PV) partial waves in the NN system, there is a chance to understand quantitatively for the first time the extensive observations performed in
many systems of parity-violating phenomena in nuclei~\cite{Adelberger85a}. Nuclear parity violation is linearly sensitive to  small components of the nuclear wavefunction since successive shell model  levels alternate in parity, and parity-odd operators directly connect adjacent shells~\cite{Adelberger85b}. Ideas from quantum chaos~\cite{Zelevinsky96} and nuclear statistical spectroscopy have been used to analyze parity violation in neutron reactions in heavy nuclei in terms of the effective isovector and isoscalar weak NN interaction, and knowledge of PV in the NN system would allow a quantitative test of the predictive power of these ideas~\cite{Bowman93,Tomsovic00}. The matrix elements for weak NN interactions in nuclei also bear many similarities to the types of matrix elements that must be calculated to interpret limits on neutrino masses from double beta decay searches~\cite{Prezeau03}.  

3) In atoms, the effect of NN parity violation was seen for the first time in $^{133}$Cs~\cite{Wood97} through its contribution to the anapole moment of the nucleus, which is an axial vector coupling of the photon to the nucleus induced in heavy nuclei mainly by the PV NN interaction~\cite{Zeldovich57, Flambaum80}. Anapole moment measurements in other atoms are possible, and experiments are in progress~\cite{Aubin02}. In the heavy nuclei for which the anapole moment is a well-defined observable, the main contribution comes from PV admixtures in the nuclear ground state wave function~\cite{Haxton01a}. In electron scattering from nucleons, PV effects are sensitive to both $Z$ exchange between the electron and the quarks in the nucleon as well as the coupling of the virtual photon to the axial current from PV interactions among the quarks in the nucleon. As the precision of PV electron scattering measurements improves  in an effort to determine the strange magnetic moment of the nucleon~\cite{Beck01a, Beck01b, Beck03, Beise05}, it may  be important to know enough about the weak NN interaction to reliably extract the information of interest.

4) The NN weak interaction is also the only practical way to study quark-quark neutral currents at low energy. The neutral weak current conserves quark flavor to high accuracy in the standard electroweak model (due to the GIM mechanism). It is not seen at all in the well-studied strangeness-changing nonleptonic weak decays. We know nothing experimentally about how QCD modifies weak neutral currents.

There are theoretical difficulties in trying to relate the underlying electroweak currents to low-energy observables in the strongly interacting regime of QCD. One expects the strong repulsion in the NN interaction to keep the nucleons sufficiently separated for a direct exchange of $W$ and $Z$ bosons between quarks to represent an accurate dynamical mechanism. If one knew weak NN couplings from experiment, they could be used to interpret parity violation effects in nuclei.  The current approach is to split the problem into two parts. The first step maps QCD to an effective theory expressed in terms of the important degrees of freedom of low energy QCD, mesons and nucleons. In this process, the effects of quark-quark weak currents appear as parity-violating meson-nucleon couplings~\cite{Desplanques80a}. The second step uses this effective theory to calculate electroweak effects in the NN interaction in terms of weak couplings.  The couplings themselves also become challenging targets for calculation from the Standard Model. 

\subsection{Theoretical Description}\label{sec:NNThy}

In the work of Desplanques, Donoghue, and Holstein (DDH)~\cite{Desplanques80a}, the authors used a valence quark model in combination with SU(6) symmetry relations and data on hyperon decays to produce a range of predictions for effective PV meson-nucleon couplings from the SM. At low energy, the weak
interaction between nucleons in this approach is parameterized by the weak pion coupling constant $f_\pi$, and six other meson coupling denoted as $h^{0}_{\rho}$, $h^{1}_{\rho}$, $h^{'1}_{\rho}$, $h^{2}_{\rho}$, $h^{0}_{\omega}$, and  $h^{1}_{\omega}$, where the subscript denotes the
exchange meson and the superscript indicates the isospin change.  Due to uncertainties in the effects of strong QCD, the range of predictions is rather broad. For the weak pion coupling, neutral currents  should play a dominant role. Another strategy is to perform a systematic analysis of the weak NN interaction using an EFT approach  to classify the interaction in a manner that is consistent with the symmetries of QCD and does not assume any specific dynamical mechanism. Such an EFT approach has recently appeared~\cite{Zhu05}.  Preparations have also been made for an eventual calculation of the weak NN interaction vertices using lattice gauge theory in the partially quenched approximation~\cite{Beane02}. 

At the low energies accessible with cold neutrons with $k_nR_{strong}<<1$, parity-odd effects in the two-nucleon system can be parameterized by the five independent amplitudes for $S-P$ transitions involving the following nucleons and isospin exchanges: $^1S_0  \rightarrow{^3P_0}$ (p-p, p-n, n-n, $\Delta I=0, 1, 2$), $^3S_1\rightarrow
{^1P_1}$ (n-p, $\Delta I=0$), and $^3S_1 \rightarrow {^3P_1}$ (n-p, $\Delta I=1$). Thus, from the point of view of a phenomenological description of the weak NN interaction, at least five independent experiments are required. The PV longitudinal analyzing power in p-p scattering, which determines a linear combination of the $^1S_0 \rightarrow  {^3P_0}$ amplitudes, has been measured at 15\,MeV and 45\,MeV in several experiments with consistent results~\cite{Potter74,Balzer80,Kistryn87,Eversheim91}
and remains the only nonzero observation of parity violation in the pure NN system dominated by p-waves.

Parity violation in few-nucleon systems should be cleanly interpretable in terms of the NN weak interaction due to recent theoretical and computational advances for the strong interaction in few nucleon systems~\cite{Pieper01}. Weak effects can be included as a perturbation. These calculations have recently been done for n-p and p-p parity violation~\cite{Carlson02,Schiavilla03,Schiavilla04} and can be done in principle for all light nuclei~\cite{Carlson04, Carlson05}. It is also possible that these microscopic calculations can be applied to systems with somewhat larger $A$, such as $^{10}$B and $^6$Li, where measurements of $P$-odd observables with low energy neutrons have reached interesting levels of sensitivity~\cite{Vesna99,Vesna03}.

\begin{figure}
\begin{center}
\includegraphics[width=6.in]{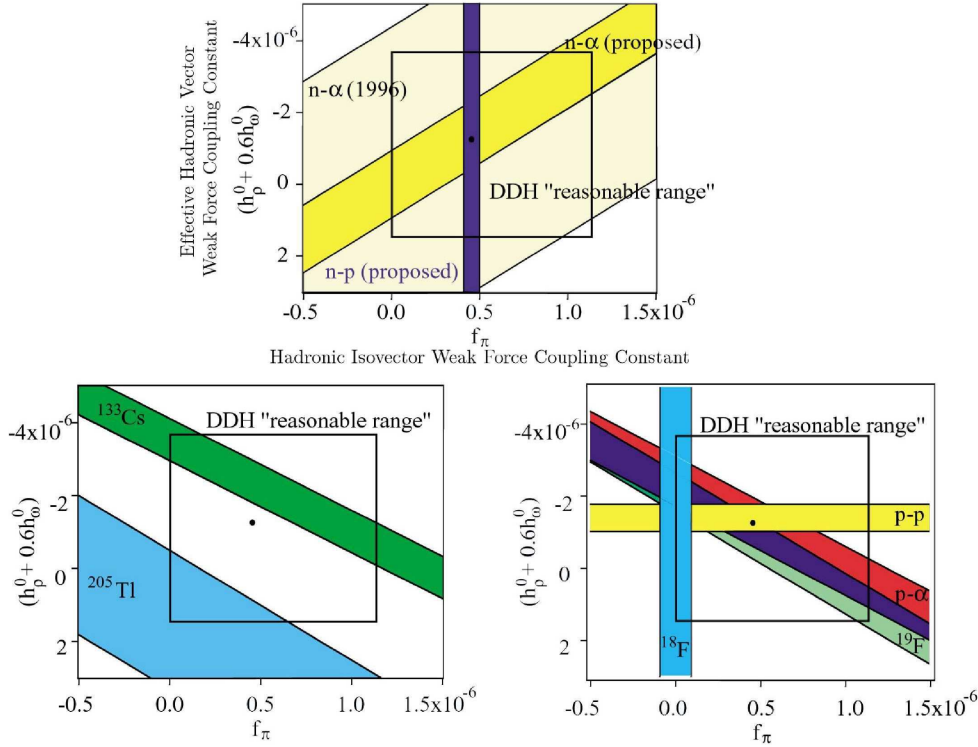}
\caption{Constraints on linear combinations of isoscalar and isovector nucleon-nucleon weak meson couplings~\cite{Haxton01b}. The bottom graphs show constraints from measurements of anapole moments in $^{133}$Cs and $^{205}$Tl and from measurements in p-p and p-$^{4}$He scattering and from $^{18}$F and $^{19}$F gamma decay. The top graph shows the anticipated constraints from proposed measurements of the $P$-odd asymmetry \NPDG\ to $5 \times 10^{-9}$ accuracy and the $P$-odd neutron spin rotation in \Hef\ to $2\times 10^{-7}$\,rad/m accuracy. In each plot, the box indicates the DDH reasonable range for the couplings.}
\label{fig:DDH}
\end{center}
\end{figure}

The longest-range part of the interaction is dominated by the weak pion-nucleon coupling constant $f_\pi$. $f_\pi$ has been calculated using QCD sum rules~\cite{Henley98, Lobov02} and in a SU(3) Skyrme model~\cite{Meissner99}, and a calculation in a nonlinear chiral quark model is in progress~\cite{Lee05}.   Measurements of the circular polarization of photons in the decay of $^{18}$F~\cite{Page87, Bini85} provide a value for $f_\pi$ that is considerably smaller than the DDH best value though still within the reasonable range. A precision atomic physics measurement of the $^{133}$Cs hyperfine structure (anapole moment) has been analyzed to give a constraint on $f_\pi$  and the combination
($h^{0}_{\rho}$+0.6$h^{0}_{\omega}$). This result would seem to favor a value for $f_\pi$ that is larger than the $^{18}$F result. Figure~\ref{fig:DDH} presents an exclusion plot that summarizes the current situation. Within the DDH one-boson exchange model of the weak NN interaction, all parity-odd amplitudes are actually products of strong and weak couplings, and changing the strong couplings to those from the Bonn potential can relax some of the observed discrepancies~\cite{Miller03}  

\subsection{Parity-Odd Neutron Spin Rotation and Capture Gamma
Asymmetries}\label{sec:CaptAymmetry}

There are a few general statements that apply to the low energy weak interactions of neutrons with low $A$ nuclei. In the absence of resonances, it can be shown that the PV helicity dependence of the total cross section vanishes if only elastic scattering is present and that both the PV neutron spin rotation and the PV helicity dependence of the total cross section with inelastic channels are constant in the limit of zero neutron energy~\cite{Stodolsky82}. These results depend only on the requirement for parity violation in an $S\rightarrow P$ transition amplitude involving two-body channels. The two practical classes of neutron experiments are PV neutron spin rotation and PV gamma asymmetries. 

The NPDGamma experiment will measure the parity-violating directional gamma ray asymmetry $A_\gamma$ in the capture of polarized neutrons on protons~\cite{Snow00, Snow03}. The unique feature of this observable is that it is sensitive to the weak pion coupling $f_\pi$, $A_\gamma=-0.11 f_\pi $~\cite{Desplanques75, Desplanques80b, Desplanques01}. The recently-commissioned beamline at LANSCE delivers pulsed cold neutrons to the apparatus, where they are polarized
by transmission through a large-volume polarized \Het\ spin filter and are transported to a liquid parahydrogen target. A resonant RF spin flipper  reverses the direction of the neutron spin on successive beam pulses using a sequence that minimizes susceptibility to some systematics.  The 2.2\,MeV gamma rays from the capture reaction are detected in an array of CsI(Tl) scintillators read out by vacuum photodiodes operated in current mode and coupled via low-noise I-V preamplifiers to transient digitizers. The current-mode CsI array possesses an intrinsic noise two orders of magnitude smaller than the shot noise from the gamma signal and has been shown in offline tests to possess no false instrumental asymmetries at the $5\times 10^{-9}$ level~\cite{Gericke05}. The pulsed beam enables the neutron energy to be determined by time-of-flight, which is an important advantage for diagnosing and reducing many types of systematic uncertainty. This apparatus has been used to conduct measurements of parity violation in several medium and heavy nuclei~\cite{Mitchell04}.

Another experiment in preparation is a search for parity violation in neutron spin rotation in liquid \Hef~\cite{Bass05}. A transverse rotation of the neutron spin vector about its momentum manifestly violates parity~\cite{Michel64} and can be viewed from a neutron optical point of view as due to a
helicity-dependent neutron index of refraction. For \Hef, the calculated PV neutron spin rotation in terms of weak couplings is~\cite{Dmitriev83}
\begin{eqnarray}
    \label{eq:Hefspinrot}
\phi =(0.97f_\pi+0.32h^{0}_{\rho}-0.11h^{1}_{\rho}+0.22h^{0}_{\omega}-0.22h^{1}_{\omega}) \times 10^{-6} \,\text{rad/m}
\end{eqnarray}
To measure the small parity-odd rotation, a neutron polarizer-analyzer pair is used to transmit only the component of a polarized beam that rotates as it traverses the target.  The challenge is to distinguish small PV rotations from rotations that arise from residual magnetic fields. The first measurement achieved a sensitivity of $14\times 10^{-7}$\,rad/m at NIST~\cite{Markoff97}, and no systematic effects were seen at the $2\times10^{-7}$\,rad/m level. 

There are four plausible experiments which employ beams of cold neutrons and involve targets with $A<5$:  measurement of the PNC gamma asymmetries in \NPDG\ and in \NDTG\ and of the PNC neutron spin rotations in $^4$He and H. Successful measurements in all of these systems, in combination with existing measurements, would have a strong impact on the knowledge of the NN weak interaction~\cite{Snow05}. 

\subsection{Test of Statistical Theories for Heavy Nuclei Matrix Elements}\label{sec:NNStatThy}

One might  assume that a quantitative treatment of NN parity violation in neutron reactions with heavy nuclei would not be feasible. A low energy compound nuclear resonance expressed in terms of a Fock space basis in a shell model might possess a million components with essentially unknown coefficients, and the calculation of a parity-odd effect would involve a matrix element between such a state and another equally complicated state. However, one can imagine a theoretical approach which exploits the large number of essentially unknown coefficients in such complicated states. If we assume
that it is possible to treat these components as random variables, one can devise statistical techniques to calculate, not the value of a particular parity-odd observable, but the width of the distribution of expected values. A similar strategy has been used to understand other properties of complicated compound nuclear states. The distribution of energy spacings has been known for a long time to obey a Porter-Thomas distribution~\cite{Porter65} in agreement with the predictions of random matrix theory, and statistical approaches have been used to understand isospin violation in heavy nuclei~\cite{Harney86}.  Statistical analyses have been applied to an extensive series of measurements of parity violation in heavy nuclei  performed mainly at Dubna, KEK, and LANSCE~\cite{Mitchell99, Mitchell01}.

The TRIPLE collaboration at LANSCE measured 75 statistically significant PV asymmetries in several compound nuclear resonances in heavy nuclei. In the case of parity violation in compound resonances in neutron-nucleus reactions there are amplification mechanisms which can enhance parity-odd observables by factors as large as $10^{5}$. These amplification mechanisms, which are interesting phenomena in themselves, depend in an essential way on the complexity of the states involved and the reaction dynamics. Part of the amplification comes from the decrease in the spacing between levels as the number of nucleons increases, which brings opposite parity states closer together and increases their weak mixing amplitudes~\cite{French88a,French88b}, and for low energy neutron reactions in heavy nuclei it leads to a generic amplification of order $10^{2}$ in parity-odd amplitudes. In addition, for low energy neutron-nucleus interactions the resonances are mainly $l=0$ and $l=1$ with the scattering amplitudes in s-wave resonances larger than for p-wave
resonances by a factor of order $10^{2}$ to $10^{3}$. At an energy close to a p-wave resonance, the weak interaction mixes in an s-wave component that is typically much larger, and this factor also amplifies the asymmetry. These amplification mechanisms were predicted theoretically~\cite{Sushkov80} before they were measured~\cite{Alfimenkov82,Alfimenkov83}.

A basic tenet of the statistical approach is that there should be, on average, equal numbers of negative and positive PV asymmetries in a given nucleus. This appears to be satisfied in all nuclei measured so far except for $^{232}$Th; below $E=250$\,eV all ten of the parity-odd asymmetries in this nucleus have the same sign. At present this result is ascribed to some poorly understood nuclear structure effect specific to $^{232}$Th. Its observation illustrates the potential for the use of NN parity violation to discover unsuspected  nuclear structure effects in complicated many body systems. In another recent example, a series of experiments that first observed parity-odd neutron spin rotation in natural Pb~\cite{Heckel82} and subsequent experiments on spin rotation in separated lead isotopes~\cite{Bolotsky96} led to a prediction of the first-known subthreshold p-wave resonance in neutron-nucleus interactions in the isotope with the large parity violation~\cite{Lobov00}. Although this isotope was thought to be $^{204}$Pb according to the results of a spin rotation experiment at HMI~\cite{Golub02}, a subsequent neutron radiative capture experiment seems to indicate that $^{207}$Pb possesses the subthreshold p-wave resonance~\cite{Andrzejewski04}.   

The values of the weak matrix elements determined by the statistical analysis varied in the range $0.5$\,meV to $3.0$\,meV, in rough agreement with theory. The accuracy of this analysis was improved through new measurements of the required spectroscopic information on compound nuclear levels~\cite{Corvi04}, and more spectroscopic information on the appropriate nuclei would be helpful.  Theoretical calculations to use this data to extract the weak isoscalar and isovector couplings in $^{238}$U obtain results in qualitative agreement with DDH expectations~\cite{Tomsovic00}. If the weak NN couplings were known, we should be able to use the results to see if there is any evidence for nuclear medium effects, which in principle should exist. Measurements of the mean square weak matrix elements in lower A regions would obviously be helpful.  

\subsubsection{Parity-odd and Time Reversal-odd Correlations in Neutron-induced Ternary Fission}\label{sec:TernFiss}

Another example where symmetry violation in neutron-induced reactions has led to progress in the understanding of many-body nuclear dynamics is fission. $P$-odd effects in binary fission induced by polarized neutron capture have been observed for a long time~\cite{Danilyan77,Vodennikov78}.
Although one might expect that a treatment of parity violation in nuclear fission
would be even more difficult than for compound nuclear resonances in heavy nuclei, there is a compelling understanding of parity-odd asymmetries observed in fission after capture by polarized
neutrons~\cite{Danilyan77,Vodennikov78,Shuskov82} based on interference of amplitudes of opposite parity from parity doublets in the relatively small number of open tunneling channels through the cold pear-shaped transition state.  Since this interference occurs among these small number of fission channels in the initial state near the saddle point, it can survive the inevitable averaging over the enormous number of final states later produced in the rupture.

In the case of ternary fission, where a third light charged particle (usually an alpha) is emitted in addition to the two main fragments, recent parity violation measurements have given support to a
specific mechanism for the emission of the ternary particle~\cite{Bunakov02a}. Consider two generic mechanisms: the simultaneous emission of the three
particles (three-body compound nucleus decay) and ``double neck rupture'' in which the ternary particle is emitted after the first rupture of the neck from its remnants. In the first case, because all three objects originate from the same system where the dominant parity violation comes from the mixing of opposite parity compound nuclear states, one expects all of the PV asymmetries in various channels to be about the same size. In the second case, however, the mechanism for the emission of the
ternary particle does not possess the same intermediate states that are known to exist in binary fission, and upon averaging over the large number of fragment states one would expect the parity-odd correlations that involve the ternary particle to be much smaller. This is what is observed experimentally in $^{233}$U~\cite{Petrov89,Belozerov91,Goennenwein94,Jesinger02,Koetzle00}. Furthermore, the parity-odd asymmetries of the two large fragments were seen to be independent of the energy of the ternary particle. Since different ternary particle energies are presumably coming from different Coulomb repulsion effects from different shapes of the neck, this independence would also seem to indicate that the parity-odd asymmetry is established before the scission process.

In ternary fission one can also look for a formally $T$-odd triple correlation between the momenta of the light fragment and ternary particle and the neutron spin. This measurement has recently been done~\cite{Jesinger99,Jesinger00} and a large nonzero effect of order $10^{-3}$ was seen in both $^{233}$U and $^{235}$U for both alphas and tritons as the ternary products. It is believed that in the ternary fission system this correlation is due to a final state effect and not to a fundamental source of $T$ violation. The fact that the size of the observed triple correlation  depends on the ternary particle energy also suggests that a final state effect is responsible. One model~\cite{Bunakov02b,Bunakov02c,Kadmensky02} can reproduce the order of magnitude of the effect if one assumes that the projection of the orbital angular momentum of the recoiling ternary particle changes the spin projections and therefore the level densities of the larger fragments. If the emission probabilities of the ternary particle are proportional to these level densities and the angular momentum of the initial system is correlated with the neutron polarization, this mechanism can generate a nonzero triple correlation. Semiclassically this can be viewed in terms of the Coriolis interaction of the ternary particle with the rotating compound nucleus~\cite{Bunakov03}. Other recent work asserts that the large $T$-odd correlation in ternary fission is evidence for the simultaneous production of the three particles~\cite{Barabanov03a}. Future work will attempt to confirm this mechanism in plutonium.

\section{LOW ENERY QCD TESTS}\label{sec:nAScattering}

One of the long-term goals of strong interaction physics is to see how the properties of nucleons and nuclei follow from QCD. For nuclei the first step in this process is to see if one can start from QCD and calculate the well-measured NN strong interaction scattering amplitudes and the properties of the deuteron. Over the last decade a number of theoretical developments have started to show the outlines of how this connection between QCD and nuclear physics can be made. In this section, we discuss some of the theoretical developments in few nucleon systems along with the several precision scattering length experiments. The status of two fundamental properties of the neutron, its polarizability and the neutron-electron scattering length, are also discussed.

\subsection{Theoretical Developments in Few Nucleon Systems and the
Connection to QCD}\label{sec:3body}

Based on a suggestion by Weinberg~\cite{Weinberg79}, one strategy to develop an effective field theory~\cite{Georgi93} for QCD that is valid in the low energy limit relevant for nuclei and incorporates the most important low energy symmetry of QCD is through chiral symmetry. It is recognized that this alone is not enough because some of the important energy scales of nuclear physics, such as the deuteron binding energy and its correspondingly large low energy scattering lengths, seem to be the result of a delicate cancelation between competing effects which will need more than chiral symmetry alone to understand. Recent efforts to understand the emergence of smaller energy scales in nuclear physics not set by chiral dynamics, such as the deuteron binding energy, have led to interesting suggestions that the low energy limit of QCD is not described by the usual  renormalization group fixed point but rather is close to a limit cycle which can be reached by a small tuning of the values of the current quark masses in the Lagrangian~\cite{Braaten03,Braaten06}. 

Recently, significant insight into certain features of few-nucleon systems has come from the EFT approach based on the chiral symmetry of QCD~\cite{Bedaque98,Hammer99,Bedaque02}. The value of the EFT approach is that it is a well-defined field theoretical procedure for the systematic construction of a low energy Lagrangian consistent with the symmetries of QCD. To a given level of accuracy the Lagrangian contains all possible terms consistent with symmetries with a number of arbitrary coefficients which, once fixed by experiment, can be used to calculate other observables. EFTs based on the chiral symmetry of QCD have been used to develop an understanding for the relative sizes of many quantities in nuclear physics, such as that of nuclear N-body forces~\cite{Friar01} and in particular the nuclear 3-body force (3N), which is the subject of much activity. Although it is well understood that 3N forces must exist with a weaker strength and shorter range than the NN force, little else is known.

EFT has been used to solve the two and three nucleon problems with short-range interactions in a systematic expansion of the small momentum region set by $kb \leq 1$, where $k$ is the momentum transfer and $b$ is the scattering length~\cite{Bedaque02, Beane02}. For the two-body system,
EFT is equivalent to effective range theory and reproduces its well-known results for NN forces~\cite{Kolck98,Kaplan98,Gegelia98}. The chiral EFT expansion does not require the introduction of an operator corresponding to a 3N force until next-to-next-to leading order in the expansion, and at this order it requires only two low energy constants~\cite{Epelbaum01,Epelbaum02}. There have also been significant advances in other approaches to the computation of the properties of few-body nuclei with modern potentials~\cite{Carlson98} such as the AV18 potential~\cite{Wiringa95,Pieper01}, which includes electromagnetic terms and terms to account for charge-independence breaking and charge symmetry breaking.

All of these developments show that precision measurements of low energy strong interaction properties, such as the zero energy scattering lengths and electromagnetic properties of small $A$ nuclei, are becoming more important for strong interaction physics both as precise data that can be used to fix parameters in the EFT expansion and also as new targets for theoretical prediction. It is possible now to envision the accurate calculation of low energy neutron scattering lengths for systems with $A>2$. 

\subsection{Precision Scattering Length Measurements Using Interferometric Methods}\label{sec:ScatLeng}

In parallel with these theoretical developments two interferometric methods have been perfected to allow high precision measurements of n-A scattering lengths. One is neutron interferometry using diffraction from perfect silicon crystals~\cite{Rauch00}, which measures the coherent scattering length. The other is pseudomagnetic precession of polarized neutrons in a polarized nuclear target, which measures the incoherent scattering length. Together these two measurements can be used to
determine the scattering lengths in both channels.

Neutron interferometry  can be used to measure the phase shift caused by neutron propagation in the optical potential of a medium. For a target placed in one arm of an interferometer, the phase shift is given by the expression $ \phi=b N D \lambda$, where $N$ is the target density, $D$ is
the thickness of the sample, $ \lambda$ is the neutron de Broglie wavelength, and $b$ is the coherent scattering length. Corrections to the relation between the phase shift and the coherent scattering length from multiple scattering effects in the theory of neutron optics are small and calculable~\cite{Sears85}. High absolute accuracy in the determination of $ N$,$ D$, and $\lambda$ are required but possible at the $10^{-4}$ level.
Recent results from measurements at the NIST Neutron Interferometry and Optics Facility yielded the coherent scattering lengths $b_{np}=(-3.738\pm 0.002)$\,fm, $b_{nd}=(6.665\pm0.004)$\,fm~\cite{Black03,Schoen03}, and $b_{n^3He}=(5.857 \pm 0.007)$\,fm~\cite{Huffman04} (see also~\cite{Ketter05}). They showed that almost all existing theoretical calculations of the n-d and n-\Het\  coherent scattering lengths are in disagreement with experiment and that the accuracy of present measurements is sensitive to such effects as nuclear three-body forces and charge symmetry-breaking~\cite{Hofmann03,Witala03}. 

If the neutron-nucleus interaction is spin dependent, a polarized neutron moving through a polarized medium possesses a contribution to the forward scattering amplitude proportional to $(b_{+}-b_{-}){\vec \sigma_{n}} \cdot {\vec I}$ where $b_{+}$ and $b_{-}$ are the scattering lengths in the two channels, $\vec \sigma_{n}$ is the neutron spin, and $\vec I$ is the nuclear polarization. The angle of the polarization of a neutron initially polarized normal to the target polarization precesses as it moves through the medium. This phenomenon is referred to as nuclear pseudomagnetic precession~\cite{Abragam73} and has been used to measure scattering length differences in many nuclei. Recently, a high-precision measurement of this precession angle was performed in polarized \Het\ using a neutron spin-echo spectrometer at the ILL~\cite{Zimmer02}. A new experiment to determine the spin-dependence of the n-d scattering length is in preparation at PSI~\cite{Brandt04}.

In combination with the n-d coherent scattering length determined by neutron interferometry, the PSI experiment should determine both n-d scattering lengths to $10^{-3}$ accuracy. The $^2$S$_{1/2}$ scattering length in the n-d system is especially interesting. The quartet s-wave scattering length ($^4$S$_{3/2}$) can be unambiguously determined from current theory.  Because the three nucleons in this channel exist in a spin-symmetric state, and hence have an antisymmetric space-isospin wavefunction, the scattering in this state is completely determined by the long range part of the triplet s-wave NN interaction in the n-p channel, i.e. by n-p scattering and the properties of the deuteron. By contrast the Pauli principle does not deter the doublet channel from exploring the shorter-range components of the NN interaction, where 3N forces should appear.

Perhaps the single most interesting scattering length to measure is the neutron-neutron scattering length $b_{nn}$. No direct measurements exist. An experiment to determine $b_{nn}$ by viewing a high-density neutron gas near the core of a pulsed reactor that produces an extremely high instantaneous neutron density and measuring a quadratic dependence of the neutron fluence on source power is currently being designed~\cite{Furman02, Mitchell05}. An experiment to let the neutrons in an extracted  beam scatter from each other has been considered~\cite{Pokotolovskii93}.  A precision (a few percent) measurement of $b_{nn}$ would be valuable, because it could give new information on the size of charge symmetry breaking effects of the strong interaction. Charge symmetry is the n-p interchange symmetry, a discrete subset of isospin symmetry corresponding to a $180$ degree rotation in isospin space.  A calculation of $b_{nn}$ using low energy effective field theories of QCD  would seem to be out of reach at present since there are so few constraints on the nature of charge symmetry breaking in the strong interaction from other systems.   EFT analyses to extract $b_{nn}$ from neutron-neutron final state effects in few body reactions are possible, and an EFT analysis of the $\pi^{-}d \to nn\gamma$ reaction to extract $b_{nn}$ has recently appeared~\cite{Gardestig06}.

It is interesting that both the n-d and n-\Het\ scattering lengths are now of interest as input for few body calculations in other systems. The n-d coherent scattering length has been used recently to improve the precision of the determination of low energy constants in a chiral effective field theory for few nucleon systems~\cite{Epelbaum06}. The improvements in the n-\Het\ scattering lengths are important input for a recent EFT calculation of the small cross section (but relevant in astrophysics) for the process $n+{^{3}\mathrm{He}}  \to {^{4}\mathrm{He}}  +\gamma$~\cite{Park06} that was previously measured at the ILL~\cite{Wolfs89}. These EFT calculations precisely reproduce radiative processes in few body systems such as the $n+p \to {\mathrm D}+\gamma$ total cross section~\cite{Cokinos77}. A number of precise measurements have been performed on radiative processes in neutron capture on few body nuclei such as the total cross section for $n+{\mathrm D} \to {^{3}\mathrm{H}} +\gamma$ and $n+{^{3}\mathrm{He}} \to {^{4}\mathrm{He}}  +\gamma$ and the circular polarization in $n+p \to {\mathrm D}+\gamma$~\cite{Bazhenov92} and $n+{\mathrm{D}}  \to {^{3}\mathrm{H}} +\gamma$~\cite{Konijnenberg88}, and there is an upper bound on the gamma asymmetry in polarized neutron capture on polarized protons $\vec{n}+\vec{p} \to {\mathrm{D}}+\gamma$~\cite{Muller00}. As an example of the present level of sophistication of the theoretical calculations in this field one can see in where two and even three-body electromagnetic current operators are needed to reproduce the measurements of radiative capture in the n-D system~\cite{Marcucci05}. EFT calculations, which are in principle traceable to QCD, can now be performed for these and other low energy observables, and it would be useful to perform more precise measurements in these systems to test the new predictive power of these approaches.       
 
\subsection{Neutron-electron Interaction}\label{sec:NeInteraction}

The neutrality of the neutron is not a consequence of the minimal 3-generation Standard Model, and in principle it can possess a nonzero electric charge proportional to the difference in (conserved) lepton numbers of different generations~\cite{Foot93}.    The stringent upper limit on the electric charge of the neutron from laboratory experiments~\cite{Baumann88}, $q_{n}<10^{-21} e$, strongly favors SM extensions, such as Majorana neutrino mass terms, which break family lepton number symmetries. One can imagine improving this direct limit using ultracold neutrons~\cite{Borisov88} although indirect limits from astrophysical arguments are much more stringent. There is also a stringent limit on a possible magnetic monopole charge of the neutron~\cite{Finkelstein86}.

Although the neutron has a net zero electric charge, it is composed of charged quarks which possess a nontrivial radial charge distribution. This distribution produces a nonzero value of the neutron mean-square charge radius $\langle r{_n}{^2}\rangle$. To first order in the electromagnetic coupling $\alpha$, the neutron-electron scattering length
\begin{equation}
a_{ne}={2 \alpha m_{n}c \over \hbar} {dG_{eN} \over dq^{2}}
\end{equation}
is proportional to the slope of the electric form factor of the neutron, $G_{eN}$,  in the $q^{2} \to 0$ limit, where $q$ is the momentum transfer. For the proton (neutron), this limiting value for the electric form factor is normalized to one (zero). This leads to $G_{eN}(-q^{2})\to {1 \over 6}\langle r_{n}^{2}\rangle q^{2}$, where $-q^{2}={\vec q}\,^{2}$ is the four-momentum transfer. Although defined for arbitrary $q^{2}$,  in the Breit frame $G_{eN}$ has an interpretation as the spatial Fourier transform of the charge distribution of the neutron~\cite{Isgur99}.

The sign of this slope, or equivalently the sign of the charge radius, has physical significance. For  the neutron one expects a negative charge radius from its virtual pion cloud~\cite{Hand63}.  From the QCD point of view the neutron charge radius is especially interesting since it is more sensitive to sea quark contributions than the proton charge radius, which has a large valence quark contribution.   With the advent of improved lattice gauge theory calculations of nucleon properties~\cite{Tang03}  and the potential to use chiral extrapolation procedures to ensure proper treatment of the nonanalytic chiral corrections~\cite{Leinweber03}, it is possible that the neutron-electron scattering length may be calculable in the near future directly from QCD. Given the recent improvements in the measurement of the neutron electric form factor at larger momentum transfers, it is becoming more important to obtain a reliable value for $a_{ne}$. 

There are two clusters of values in  $a_{ne}$ measurements. One set comes from  measurements of the asymmetric angular distribution of neutron scattering in noble gases, $a_{ne}=(-1.33 \pm 0.03) \times 10^{-3}$\,fm~\cite{Krohn66,Krohn73}, and the total cross section of lead, $a_{ne}=(-1.33 \pm 0.03) \times 10^{-3}$\,fm~\cite{Kopecky97}. The other set comes from measurements of the total cross section of bismuth, $a_{ne}=(-1.55 \pm 0.11) \times 10^{-3}$\,fm~\cite{Alexandrov86}, and neutron diffraction from a tungsten single crystal, $a_{ne}=(-1.60 \pm 0.05) \times 10^{-3}$\,fm~\cite{Alexandrov85}. Although different contributions to these measurements from neutron resonances and interference corrections to the total cross section from coherent elastic scattering  have been discussed as possible sources for this discrepancy~\cite{Leeb93, Alexandrov96, Ignatovich99}, there seems to be no broadly accepted consensus on the explanation for the difference. Two new experiments are in preparation; one exploits dynamical diffraction in a perfect silicon crystal~\cite{Wietfeldt06}, and a second attempts to improve on the technique of scattering in noble gases~\cite{Enik03}. An experiment using Bragg reflections in perfect silicon crystals to determine $a_{ne}$ has also been proposed~\cite{Sparenberg02}. A calculation for $a_{ne}$ has been performed in a recent paper~\cite{Pineda05}.

The precision of the charge radii of the proton and the deuteron has greatly improved over the last decade from theoretical and experimental advances in electron scattering and atomic physics. The charge radius of the proton is well-determined from both electron scattering data, $\sqrt{r_{p}^{2}}=(0.895\pm 0.018)$\,fm~\cite{Sick03}, and from high precision atomic spectroscopy in hydrogen, $\sqrt{r_{p}^{2}}=(0.890\pm 0.014)$\,fm~\cite{Udem97}. The charge radius of the deuteron is also well-determined from electron scattering data $\sqrt{r_{d}^{2}}=(2.128\pm 0.011)$\,fm~\cite{Sick96}, a value consistent with theoretical calculations of deuteron structure from the deuteron wave function and the triplet n-p scattering length ($\sqrt{r_{d}^{2}}=2.131$\,fm~\cite{Klarsfeld86}), and from high precision atomic spectroscopy measurements of the 2$P$-2$S$ transition in deuterium ($\sqrt{r_{p}^{2}}=(2.133\pm 0.007)$\,fm~\cite{Pachucki94}). Atomic physics measurements of the H-D isotope shift of the $1S$-$2S$ two-photon resonance were used to derive an accurate value for the difference between the mean-square charge radii of the deuteron and proton of $r_{d}^{2}-r_{p}^{2}=(3.8212\pm 0.0015)$\,fm$^{2}$~\cite{Huber98}. Since the neutron charge radius is simply related to the proton and deuteron charge radii, it is very timely for a theoretical analysis that uses these precise values as input and predicts the neutron mean-square charge radius in an EFT analysis.  

\subsection{Neutron Polarizability}\label{sec:Polarizability}

The electric and magnetic polarizabilities of the neutron characterize how easily the neutron deforms under external electromagnetic fields.  The quarks in the neutron are confined by the strong interaction with a tension equivalent to about one ton over their distance of separation of one fermi, so the polarizabilities are very small. To measure neutron polarizability, one may exploit the electric fields accessible on the surface of heavy nuclei which polarize the neutron to give a calculable contribution to the neutron-nucleus scattering length with a linear dependence on the neutron momentum $k$.  A measurement using $^{208}$Pb  observed a term whose size and neutron energy dependence was consistent with a nonzero polarizability of $\alpha_{n}=[12.0 \pm 1.5\text{(stat)} \pm 2.0\text{(sys)}]\times 10^{-4}$\,fm$^{3}$~\cite{Schmiedmayer91}. Subsequent analyses have asserted that the data analysis is not definitive~\cite{Koester95,Aleksejeva97,Wissmann98}. Another approach using deuteron Compton scattering gave $\alpha_{n}=[8.8 \pm 2.4 \text{(stat+sys)} \pm 3.0 \text{(theo)}] \times 10^{-4}$\,fm$^{3}$~\cite{Lundin03} while quasi-free Compton scattering from the deuteron  gave $\alpha_{n}=[12.5 \pm 1.8 \text{(stat)}^{+1.1}_{-0.6} \text{(sys)} \pm 1.1 \text{(theo)}]\times 10^{-4}$\,fm$^{3}$~\cite{Kossert02}. The theoretical uncertainties come from different treatments of strong interaction effects.

QCD effective field theory is developing into a quantitative theory for the calculation of low energy nucleon properties. The lowest-order prediction of chiral perturbation theory for the neutron polarizability is $\alpha_{n}=12.2 \times 10^{-4}$\,fm$^{3}$~\cite{Bernard91}, and analysis of the extensive new Compton scattering data on the proton and deuteron using an EFT approach has extracted a neutron polarizability of $\alpha_{n}=[13.0 \pm 1.9\text{(stat)}^{+3.9}_{-1.5}\text{(sys)}]\times 10^{-4}$\,fm$^{3}$ in agreement with the chiral perturbation theory prediction~\cite{Beane05}.  Lattice gauge theory calculations of the polarizability are also improving~\cite{Christensen05}, raising the possibility of a sharp prediction for the neutron electric polarizability directly from QCD. This work should motivate further efforts to improve the neutron polarizability measurement.

\subsection{Ultrahigh Resolution Gamma Spectroscopy and Metrology with Neutrons}\label{sec:GAMS}

A rather unexpected use of low energy neutrons in nuclear physics and absolute metrology lies in the field of ultrahigh resolution gamma spectroscopy~\cite{Borner01}. The energy of a slow neutron is several orders of magnitude smaller than the few MeV energies of the gammas that are usually emitted upon neutron capture in a nucleus. As the excited nucleus emits a cascade of gammas the nucleus recoils and the subsequent gammas possess Doppler-shifted energies. The narrow lineshapes of these gammas, which can be resolved with $10^{-6}$ precision using diffraction from perfect crystals, are sensitive to the lifetimes of the nuclear states in the cascade. This technique of Gamma Ray Induced Doppler Broadening (GRID)~\cite{Borner88, Borner93} has been developed at the ILL and used to identify interesting two-phonon states in various deformed rare-earth nuclei starting with $^{168}$Er~\cite{Borner91} which are of interest for understanding the degree of collectivity in these highly-deformed states of matter. 

The very high resolution of the crystal diffraction gamma spectrometers coupled with the special knowledge of the lattice constants of the Si and Ge crystals used, which can be calibrated absolutely in terms of the meter, also make possible absolute measurements of nuclear binding energies of certain nuclei by measuring the energies of all the gammas released in the cascade. These measurements have been performed using the GAMS4 crystal spectrometer at the ILL~\cite{Kessler01}. The measurement of the very small diffraction angles of the gammas are performed with Michelson angle interferometers of nanoradian precision. This device has been used to determine the mass of the neutron to a precision of $2 \times 10^{-7}$ in atomic mass units.~\cite{Kessler99} Recent measurements of the binding energies of $^{29}$Si and $^{33}$S are in agreement with independent determinations using the cyclotron frequencies of ions in a Penning trap~\cite{Dewey06} and have been used to achieve the most precise direct test of the formula $E=mc^{2}$, which is verified at the $4 \times 10^{-7}$ level~\cite{Rainville05}.

Another unexpected application of precision neutron measurements lies in the determination of the electromagnetic fine structure constant $\alpha$. By simultaneously measuring both the wavelength and the velocity of a neutron beam, one can measure the quantity ${h \over m_{n}}=v \lambda$. Comparing with the Rydberg constant $R_{\infty}={1/2}{m_{e} \over h}c\alpha^{2}$ and expressing ${m_{e} \over h}=({m_{e} \over m_{p}})({m_{p} \over m_{n}})({m_{n} \over h})$ and noting that the mass ratios are known to very high accuracy from other measurements, one sees that a measurement of ${h \over m_{n}}$ determines $\alpha$. This ratio was measured by selecting a wavelength using backscattering from a perfect silicon crystal and using polarized neutrons and a resonant neutron spin flipper to measure the velocity~\cite{Kruger95}. The result of $\alpha^{-1}=137.0360113(52)$ is in agreement with and of comparable accuracy to other measurements.

\section{NEUTRONS IN ASTROPHYSICS AND GRAVITY}\label{sec:astrophys}

This section discusses some of the ways in which neutrons and nuclear reactions involving neutrons play vital roles in several astrophysical processes. Neutrons play a decisive role in determining the element distribution in the universe. The decay rate of the neutron determines the amount of primordial \Hef\ in BBN theory, and neutron reactions in stars form most heavy nuclei beyond iron. In addition, one can use the fact that neutrons in the gravitational field of the Earth see potential differences comparable to those from the strong and electromagnetic interactions as an opportunity to search for gravitational effects on an elementary particle.

\subsection{Big Bang Nucleosynthesis}\label{sec:BBN}

Neutron decay influences the dynamics of Big Bang Nucleosynthesis (BBN) through both the size of the weak interaction couplings $g_{A}$ and $g_{V}$ and the lifetime. The couplings determine when weak interaction rates fall sufficiently below the Hubble expansion rate to cause neutrons and protons to fall out of chemical equilibrium, which occurs on the scale of a few seconds, and thus the $n/p$ ratio decreases as the neutrons decay. The lifetime determines the fraction of neutrons available as the universe cools, most of which end up in \Hef~\cite{Schramm77}, and occurs on the scale of a few minutes. The neutron lifetime remains the most uncertain nuclear parameter in cosmological models that predict the cosmic \Hef\ abundance~\cite{Lopez99, Burles99}. With the recent high-precision determination of the cosmic baryon density reported by the WMAP measurement of the microwave background~\cite{Spergel03}, there is a growing tension between the BBN prediction for the \Hef\ abundance, which is quite sensitive to the neutrino sector of the SM,  and that inferred from observation~\cite{Cyburt03}. Recall that BBN calculations predicted that the number of light neutrinos that couple to the $Z$ was about $3$ before the LEP measurements ended all doubt. The quantitative success of BBN is now routinely used to constrain various aspects of physics beyond the SM. The small size of the baryon density relative to the density required to close the universe is one of the observational cornerstones of the dark matter problem in astrophysics.  

The main concern in deviations between BBN theory and experiment remains the astronomical determinations of the \Hef\ abundance, whose systematic errors are perhaps not yet fully understood. As the neutron lifetime measurements improve, other neutron-induced reactions in the early universe, such as the $n+p \rightarrow d+\gamma$ cross section, will become more important to measure precisely. Some reactions that are difficult to measure would benefit from the application of EFT methods for calculation in the relevant energy regime, and again low energy neutron measurements will be useful to fix the EFT parameters. It is therefore likely that neutron measurements will continue to be relevant for BBN.     

\subsection{Stellar Astrophysics}\label{sec:Astro}

Nuclear reactions involving neutrons play vital roles in many astrophysical scenarios. For example, virtually all elements heavier than iron were made in environments inside stars and supernovae or more exotic environments where neutron interactions dominate the nucleosynthesis. Neutrons dominate because charged particle rates are highly suppressed by large Coulomb barriers and because astrophysical conditions favor the release of large fluxes of neutrons. In addition, many of the most interesting new astronomical results, from the latest generation of observatories to measurements of isotopic anomalies in meteorites, are providing views of the results of this nucleosynthesis with unprecedented detail and precision. Furthermore, more realistic models of astrophysical environments made possible by recent advances in computing are providing new insights into the inner workings of these objects, as well as contributing to related topics such as galactic chemical evolution and the formation and age of our solar system. 

An examination of the observed elemental abundances in the solar system, together with rudimentary nuclear physics considerations, reveals that neutron capture reactions  are essential for the origin of the elements heavier than iron~\cite{Kappeler98}.  Almost all these elements are thought to have been synthesized inside stars, supernovae, or other more exotic environments through sequences of neutron capture reactions and beta decays during the so-called slow neutron capture (``s'')~\cite{Kappeler99}  and rapid neutron capture (``r'')~\cite{Qian03}  processes. The s and r processes are each responsible for roughly half of the observed heavy element abundances. The remaining neutron-deficient isotopes that cannot be reached via neutron capture pathways are thought to have been formed in massive stars or during supernova explosions through the photodissociation (``p'') process.

In some cases, further progress in these areas is hampered by the lack of accurate rates for nuclear reactions governing stellar nucleosynthesis. Many of these astrophysical reaction rates can be determined by measuring neutron-induced cross sections in the energy range between approximately 1\,eV and 300\,keV. For about 20 radionuclides along the s-process path, the neutron-capture and $\beta$-decay time scales are roughly equal. The competition between neutron capture and $\beta$ decay occurring at these isotopes causes branches in the s-process reaction path that, if measured, could be used to directly constrain dynamical parameters of s-process models. There is very little data on the (n,$\gamma$) reaction rates for such radioactive branching points. 

Measurements of cross sections in an energy range relevant to astrophysics  have been conducted at facilities such as ORELA~\cite{Koehler01},  GELINA~\cite{Flaska04},  nTOF~\cite{Borcea03},  DANCE~\cite{Reifarth04}, and IREN~\cite{Ananev05} along with others and could also be pursued at new epithermal neutron sources such as the SNS and JSNS~\cite{Koehler02} . The intensities now suffice to conduct cross section measurements on small quantities of unstable isotopes. As the understanding of the origins and processes which lead to element formation improves, we can more effectively exploit astrophysical observations to constrain the understanding of what phenomena may lie beyond the SM.

\subsection{Gravitationally-induced Phase Shift}\label{sec:gravity}

The contribution of precision neutron measurements to gravitation are few in number but notable in conceptual impact. The equivalence principle for free neutrons has been verified at the $10^{-4}$ level~\cite{Schmiedmayer89}, and although one might do better with ultracold neutrons~\cite{Pokotilovskii94}, experiments with bulk matter are several orders of magnitude more precise.  Another connection between neutron physics and gravity is the observation of the gravitational phase shift by neutron interferometry, which was the first verification that the principles of quantum mechanics seem to apply to the gravitational potential as well as the potentials produced by other interactions~\cite{Colella75,Rauch74}.  This measurement has been performed with an accuracy at the 1\,\% level, and there is a slight disagreement between theory and experiment~\cite{Littrell97}. The source of the difference is believed to lie in gravitationally-induced distortions in the perfect crystal interferometer as it is rotated to change the relative height of the two paths in the interferometer.  There are two plans underway to improve the measurement, one to conduct the experiment with the interferometer suspended in a neutrally-buoyant fluid to eliminate possible distortions of the interferometer crystal~\cite{Kaiser06} and another to use a recently-developed Mach-Zehnder interferometer with cold neutrons~\cite{Funahashi04}.

\subsection{UCN Gravitational Bound States}\label{sec:UCNBound}

Experimental tests of gravity and searches for new long-range forces have attracted more interest in recent decades. A reanalysis~\cite{Fischbach86} of an experiment on the principle of equivalence by E\"otv\"os motivated a series of precise tests of the principle of equivalence culminating in the torsion balance experiments of the E\"ot-Wash group, which set stringent new limits to violations of the equivalence principle~\cite{Gundlach97}. Recently, speculations involving the propagation of the gravitational field into extra dimensions produce as a natural consequence a modification of the inverse square law for gravity on submillimeter scales~\cite{Sundrum99}. The current experimental limits on the existence of such forces are not stringent enough to rule out this possibility. Experiments that use smaller separations in torsion-balances~\cite{Adelberger03} and small mechanical resonators~\cite{Long03} have already set useful limits. Such experiments share some techniques with those used in the recent experimental demonstration of the Casimir effect~\cite{Lamoreaux97}.

At first glance experiments with neutrons would not seem to offer a productive technique to conduct sensitive searches for new short-range forces of gravitational strength. However, the very small polarizability of the neutron is an advantage relative to atoms, whose van der Waals attraction poses a background issue for such searches.  Although neutron beam densities are very small compared to bulk matter, the delicate control possible for neutron polarization and the weak interaction of the neutron with material media make it possible to pass the neutron through bulk samples, thus presenting an interesting opportunity to probe new forces at a distance scale that is difficult to reach with other methods. Possible experiments to probe extra dimensional gravity theories using the angular and neutron energy dependence of neutron scattering from spin zero nuclei have also been discussed~\cite{Frank04}. 
 
The recent measurements in search of gravitational bound states of ultracold neutrons are of some interest for constraints on new forces at smaller distance scales~\cite{Nesvizhevsky02, Nesvizhevsky03}. For neutron kinetic energies small compared to the neutron optical potential of a plane surface, the potential seen by a neutron moving above the plane in the gravitational field of the Earth should possess neutron bound states with energies given to good approximation by the zeroes of Airy functions, which are solutions to the Schr\"odinger equation in a linear potential. The lowest bound state of energy $1.4$\,peV hovers above the medium at a distance of order $10\,\mu$m, and any nonstandard attractive interaction of the neutron with the matter on this length scale could create another bound state. 

The experiments were designed to populate these bound states by forcing the UCN to pass through a narrow gap above a planar medium and to detect their presence by measuring the transmission of the UCN through the gap as the separation is varied. Intensity oscillations in the transmission are observed which can be fit to a model which includes the effect of the spatial extent of the bound states. The agreement of this data with theory was used to set an interesting bound on gravitational-strength forces in the nanometer range~\cite{Nesvizhevsky04, Nesvizhevsky05}. 

Other sources of constraints on new forces in this distance regime from neutron measurements come from comparisons of the results of different experiments to measure neutron-electron scattering lengths. Total cross section measurements are sensitive to the forward scattering amplitude through the optical theorem, but measurements of the angular distribution of the differential cross section depend on the momentum transfer. In the presence of a new force on the distance scale set by the momentum transfer, the neutron-electron scattering amplitude inferred from the two measurements could be different. The agreement of the neutron-electron scattering length from these two different measurements sets a competitive limit on new forces at the nanometer scale. Analysis of total neutron cross section measurements on $^{208}$Pb using eV to keV energy neutrons can also be used to set a limit for shorter distances~\cite{Leeb92}.    

\section{SUMMARY}\label{sec:Summ}

Slow neutron experiments address fundamental scientific issues in a surprisingly large range of physics subfields. The development of new types of ultracold neutron sources, pulsed and CW spallation sources, and the continued increase in the fluence of reactor beams is making possible new types of experiments and opening new scientific areas. We anticipate significant progress in the areas of neutron decay measurements and neutron EDM searches and also in the field of weak NN interactions and the measurement other low energy neutron properties. We also expect more use of neutron measurements for applications in astrophysics.  As is typical, many of the new opportunities made possible by technical developments and the emerging scientific issues were not clearly foreseen a decade ago. The steady progress over the last decade in quantitative theoretical understanding of the strong interaction is an underappreciated development with exciting applications to neutron physics.

Although still a relatively small field compared to other areas in nuclear and particle physics, the expanding scientific opportunities in fundamental neutron physics are attracting a growing number of young researchers. This growth is driving the increasing number and variety of new facilities. The diverse applications and the location of neutron physics at facilities whose main purpose is generally neutron scattering have combined to obscure somewhat its accomplishments. We hope that the reader of this review has gained an appreciation for the breadth of activity in the field.

\section{ACKNOWLEDGMENTS}\label{sec:Ack}

We would like to thank Torsten Soldner of the ILL and Yasuhiro Masuda of KEK for their assistance in supplying some of the parameters for neutron facilities, and Scott Dewey for his careful reading of the manuscript. W.M.S. gratefully acknowledges support from the National Science Foundation, the Department of Energy, and the Indiana 21st Century Fund. J.S.N. acknowledges the support of the NIST Physics Laboratory and Center for Neutron Research.

\newpage
%\bibliographystyle{apsrev}
%\bibliographystyle{ieeetr}
%\bibliography{AnnRevref}

%\addcontentsline{toc}{section}{References}
%\bibliographystyle{prsty}
%\bibliographystyle{plain}

\end{document}